\documentclass[preprint,12pt]{elsarticle} 

\usepackage{graphicx}
\usepackage{float}
\usepackage{subcaption}
\usepackage{changepage}
\usepackage{booktabs}
\usepackage{algorithm}
\usepackage{algorithmic}
\usepackage{enumerate}
\usepackage[T1]{fontenc}
\usepackage{amssymb}
\usepackage{pdflscape}
\usepackage{afterpage}
\usepackage{capt-of}
\usepackage{multirow}
\usepackage{array}
\usepackage[table]{xcolor}
\usepackage{algorithm}
\usepackage{algorithmic}
\usepackage{enumerate}
\usepackage{setspace}
\usepackage{todonotes}
\usepackage{amsmath}
\usepackage{comment}
\usepackage{changepage}
\usepackage{upgreek}
\usepackage{siunitx}
\usepackage{url}
\usepackage[normalem]{ulem}
\usepackage{textcomp}
\usepackage{atbegshi,ifthen,listings,tikz}
\tikzstyle{highlighter} = [
  gray!25,
  line width = 0.6\baselineskip,
]

\newcounter{highlight}[page]

\AtBeginShipout{\AtBeginShipoutUpperLeft{\ifthenelse{\value{highlight} > 0}{\tikz[remember picture, overlay]{\foreach \stroke in {1,...,\arabic{highlight}} \draw[highlighter] (begin highlight \stroke) -- (end highlight \stroke);}}{}}}

\usepackage{amssymb} 

\newcommand{\systemname}{MIDAS}

\newcommand{\SystemName}{MIDAS}

\newcommand\cparagraph[1]{\vspace{1.0mm}\noindent\textbf{#1:}}
\def\isfinal{1}
\newcommand{\wipcomment}[2]{\if\isfinal0{\color{#1}{#2}}\else{#2}\fi}

\def\isfinal{1}
\newcommand{\rev}[1]{\if \isfinal0{\textcolor{blue}{#1}}\else {#1}\fi}

\journal{.}
\usepackage{enumitem} 
\setlist{nolistsep}

\let\OLDthebibliography\thebibliography
\renewcommand\thebibliography[1]{
  \OLDthebibliography{#1}
  \setlength{\parskip}{0pt}
  \setlength{\itemsep}{0pt plus 0.3ex}
}

\begin{document}

\begin{frontmatter}


\title{Thermal Dissipation Resulting from Everyday Interactions as a Sensing Modality-The \SystemName{} Touch}

\author[rvt1]{Farooq Dar}
\ead{farooq.dar@ut.ee}

\author[rvt1]{Hilary Emenike}
\ead{hilary.emenike@ut.ee}

\author[rvt1]{Zhigang Yin}
\ead{zhigang.yin@ut.ee}

\author[rvt1]{Mohan Liyanage}
\ead{mohan.liyanage@ut.ee}

\author[rvt1]{Rajesh Sharma}
\ead{rajesh.sharma@ut.ee}

\author[rvt2]{Agustin Zuniga}
\ead{agustin.zuniga@helsinki.fi}

\author[rvt2]{Mohammad A. Hoque}
\ead{mohammad.hoque@helsinki.fi}

\author[rvt3,rvt4]{Marko Radeta}
\ead{marko@wave-labs.org}

\author[rvt2]{Petteri Nurmi}
\ead{petteri.nurmi@helsinki.fi}

\author[rvt1]{Huber Flores}
\ead{huber.flores@ut.ee}

\address[rvt1]{Institute of Computer Science, University of Tartu, Estonia,}
\address[rvt2]{Department of Computer Science, University of Helsinki, Finland,}
\address[rvt3]{Wave Labs, MARE/ARDITI, University of Madeira, Portugal,}
\address[rvt4]{Department of Astronomy, Faculty of
Mathematics, University of Belgrade, Serbia, \\
firstname.lastname@\{ut.ee, helsinki.fi, mare-centre.pt\}}

\begin{abstract}

We contribute \SystemName{} as a novel sensing solution for characterizing everyday objects using thermal dissipation. \SystemName{} takes advantage of the fact that anytime a person touches an object it results in heat transfer. 
By capturing and modeling the dissipation of the transferred heat, e.g., through the decrease in the captured thermal radiation, \SystemName{} can characterize the object and determine its material. We validate \SystemName{} through extensive empirical benchmarks and demonstrate that \SystemName{} offers an innovative sensing modality that can recognize a wide range of materials -- with up to $83\%$ accuracy -- and generalize to variations in the people interacting with objects. We also demonstrate that \SystemName{} can detect thermal dissipation through objects, up to \SI{2}{mm} thickness, and support analysis of multiple objects that are interacted with.

\end{abstract}

 
\end{frontmatter}

\section{Introduction}

Every day humans touch numerous objects, ranging from their personal possessions to home appliances, food, clothing, and many other objects~\cite{zuccotti2015every}. Capturing information about the objects people interact with has the potential to offer rich insights into human behaviour~\cite{klatzky1985identifying}. Such ranges from simple everyday activity monitoring to more complex applications, being dietary monitoring~\cite{bi2018auracle} or detection of household practices~\cite{babaei2015household}. Unfortunately, capturing such information is fraught with difficulty as current solutions suffer from some significant limitations. Specifically, contact-based approaches, such as RFID, either require instrumenting all objects or keeping the sensing device in close contact with the object for a sufficiently long period~\cite{ha2018learning, wang2017tagscan}. Non-contact based solutions, such as image-based object recognition, in turn, have limited discriminatory power and are sensitive to the conditions of the operating environment~\cite{yeo2017specam}.
For example, image-based recognition is prone to changes in illumination, camera angle, and picture resolution~\cite{schwartz2019recognizing}.

This paper develops \SystemName{} as a novel sensing solution for characterizing everyday objects using the thermal dissipation resulting from human touch. \SystemName{} exploits the fact that anytime a person touches an object it results in heat transfer. Over time, the transferred heat dissipates from the object as the object attempts to reach thermal equilibrium with the surrounding environment. The rate of this dissipation depends on the material characteristics of the object. \SystemName{} captures the heat transfer and the ensuing dissipation and uses these to model and characterize the materials of the objects that the user interacts with. \SystemName{} captures the changes in heat using a commercial-off-the-shelf (COTS) thermal camera. Thermal cameras can operate without requiring contact and they are robust to illumination conditions and the overall capture context, thus overcoming the key limitations of current techniques. Thermal imaging is also increasingly feasible thanks to the increase of thermal cameras in production. For example, several Caterpillar smartphones integrate forward looking infrared (FLIR) thermal cameras and there are affordable external thermal cameras that can be connected with smartphones. These cameras provide opportunity to capture thermal transfer and to monitor the speed of the consequent heat dissipation to offer insights of daily interactions with objects and the materials of these objects. 


We validate \SystemName{} through rigorous experiments that consider $14$ different everyday objects that cover the most common materials used in manufactured products. As part of the experiments, we also demonstrate that \SystemName{} generalizes to human temperature variations by considering the robustness of thermal dissipation characteristics of objects with $18$ different individuals. Our results indicate that human-emitted radiation can be used to characterize different materials and that such characterization is robust against variations in individuals and the way they interact with objects themselves. \SystemName{} can determine the correct material with up to $83\%$ accuracy, a $16\%$ improvement on a computer vision baseline that uses deep learning. 
\rev{We also demonstrate that \SystemName{} can characterize materials through other materials and support the simultaneous characterization of multiple objects.}
\SystemName{} thus offers an innovative sensing modality that can accurately and robustly characterize everyday objects and enable a broad range of innovative applications.


\subsubsection*{\textbf{Summary of Contributions}}

\begin{itemize}[noitemsep,leftmargin=*]
\item \textbf{Novel method:} We develop \SystemName{} as a novel sensing approach for characterizing materials using thermal dissipation fingerprints. We demonstrate its usefulness in material classification \rev{and show that \SystemName{} can classify different material types based on the thermal dissipation rate.} 

\item \textbf{Novel insights:} We demonstrate that current state-of-the-art techniques based on computer vision are limited and only capable of recognizing products that are not mixed with other materials. We also highlight the importance of analyzing objects' material properties to increase robustness of results. 

\item \textbf{Improved performance:} We perform rigorous benchmarks demonstrating that \SystemName{} significantly improves the classification of materials of different sizes and shapes and that is robust across when used among different persons.


\item \textbf{\rev{Detection through objects:}} \rev{We demonstrate that \SystemName{} can characterize objects through other materials (up to approximately \SI{2}{\milli\meter} thickness), which increases the practicability and applicability of \SystemName{}. We further highlight this potential by briefly discussing application scenarios that can monitor user interactions through objects or from the internal surfaces of the objects.}

\item \textbf{\rev{Characterization of multiple objects:}} \rev{We extend \SystemName{} to support characterization of multiple objects simultaneously and conduct experiments to analyze the performance of multiple object characterization. The results show that the response time is nearly constant, with increase in the amount of data or the amount of objects having only a marginal impact on response time. In terms of accuracy, the performance depends on how well different objects can be separated, which in turn depends on the constellation of objects and the resolution of the image relative to the objects. }

\end{itemize}


\section{Feasibility Assessment} \label{sec:materialselection}

We first conduct two preliminary studies to demonstrate that heat transfer from humans can be used to characterize different materials and household objects. We capture thermal radiation using COTS smartphone (CAT S60) and validate the measurements using a thermometer scanner that serves as a reference instrument. All statistically significant differences were separately validated using measurements from the reference device. We first describe the testbed setup before detailing the results.

\subsection{Testbed}
\label{ssec:testbed}

We capture thermal fingerprints using two devices: a handheld thermal imaging scanner (FLIR TG267) and a Caterpillar smartphone (CAT S60) with an integrated FLIR thermal camera. In all experiments we place the smartphone on a tripod at a distance of \SIrange{30}{35}{cm} 
from the object. We performed manual calibration on the camera after it had attained thermal equilibrium with the environment -- room temperature of \SIrange{22}{23.5}{\degreeCelsius}. 
The video was recorded with the CAT S60, while thermal reference photos were taken with the TG267 scanner. The room's ambient temperature was measured using a Netatmo weather station\footnote{\url{https://www.netatmo.com}}. Dissipation times were estimated automatically from the thermal video and validated by comparing against a ground truth obtained from a manual inspection of the video with a stopwatch. 

\subsection{Plastic Thermal Fingerprint Dissipation}

\begin{figure*}
	\centering
		\includegraphics[width=1.0\textwidth]{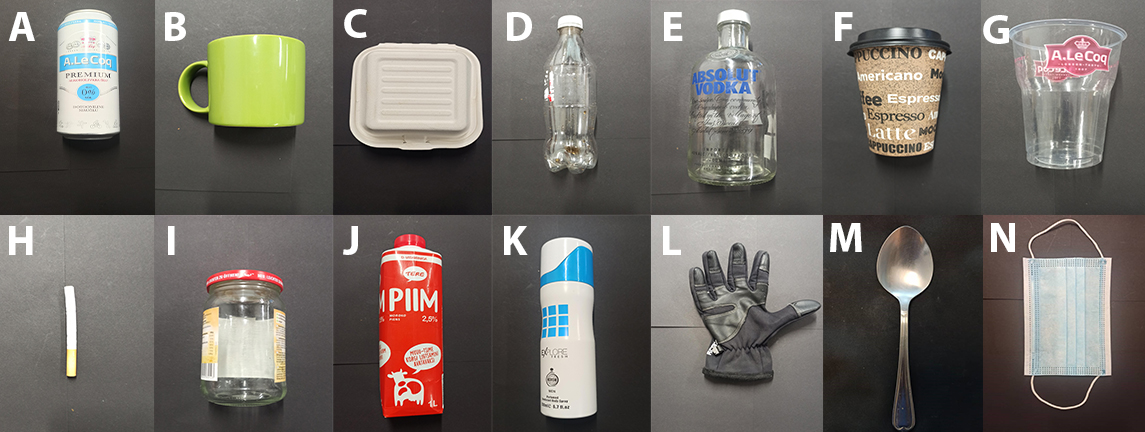}
	\caption{Selected waste materials for preliminary experiments: A (Beer Can), B (Ceramic Cup), C (Takeaway Box), D (Plastic bottle), E (Glass Bottle), F (Coffee Cup), G (Plastic Cup), H(Cigarette Butt), I (Glass Jar), J (Milk pack), K (Aerosol Can), L (Rubber glove), M (Metal spoon) and N (Face mask).}
	\label{fig:EngProto}
\end{figure*}

\cparagraph{Experimental Design} We first measure the dissipation time of a thermal fingerprint in different plastic materials and correlate the captured fingerprint with the emissivity coefficient ($\epsilon$) of the material. In this experiment we focus solely on plastics to ensure the emissivity of the materials is known. In subsequent sections, we demonstrate that our solution generalizes to other materials. We consider the most common plastics that can be found in everyday objects: LDPE (Low Density Polyethylene), HDPE (High Density Polyethylene), PP (Polypropylene), PS (Polystyrene), and PVC (Polyvinyl chloride). The material of an object is specified by its Resin Identification Code (RIC), and all materials have well known emissivity coefficients ($\epsilon$=$0.90$ - $0.97$). The tested plastic samples have identical shape and size and have been produced by an identical manufacturing process\footnote{\url{https://www.materialsampleshop.com/products/plastics-sample-set}}. This ensures the samples' differences result from inherent material properties and are not an artifact of any external differences (e.g., shape or stiffness). In the experiments, we first place the plastic sample inside a fridge with a constant temperature of \SI{5}{\degreeCelsius}, to obtain a baseline temperature for comparison. To measure different temperatures, we use a constant heat source (lamp bulb of \SI{60}{\watt}) to heat the plastic samples. The lamp is placed at a fixed distance of \SI{10}{\centi\metre} from the samples to avoid burn damage, ensuring they are exposed to sufficient amounts of thermal radiation. We consider different heating periods ($1$, $2$, $3$ and $4$ minutes) to correspond to differing initial temperatures and measure the dissipation of the thermal fingerprint. \rev{The relative changes in dissipation are highest during the initial minutes and hence periods beyond $4$ minutes were omitted.} During the experiments, ambient temperature oscillated from \SIrange{22}{24}{\degreeCelsius}. 



\cparagraph{Results} The results in Figure~\ref{fig:plastics} indicate that the dissipation of thermal fingerprints varies across the materials. The Spearman correlation~\cite{schober2018correlation}  
between dissipation time and emissivity coefficient of the materials was found statistically significant ($\rho=0.66$, p$<.05$), indicating that the dissipation characteristics indeed provide information about the material of the object.


\subsection{Other Thermal Fingerprint Dissipations}



\cparagraph{Testbed} We next demonstrate that the findings of the previous section generalize to other objects and materials by measuring the dissipation time of a thermal fingerprint on different household objects. We consider common household objects, shown in Figure~\ref{fig:EngProto}, including: a beer can (A), ceramic cup (B), takeaway box (C), plastic bottle (D), glass bottle (E), coffee cup (F), plastic cup (G), cigarette butt (H), glass jar (I), milk pack (J), aluminum aerosol can (K), rubber glove (L), steel spoon (M) and a face mask (N). \rev{We measure the dissipation of the thermal fingerprint using a thermal phone CATS60, and a certified thermometer CAT TG267. A detailed description of the apparatus is provided in Section~\ref{sec:experimentalsetup}}.
Similar to our previous experiment, we analyze the thermal dissipation time after the objects are held for $1, 2, 3$, and $4$ minutes. The average body temperature of the human subject holding the object ranged from \SIrange{35}{36}{\degreeCelsius}, and the ambient temperature was from \SIrange{22}{24}{\degreeCelsius}.

\begin{figure*}[!t]
\centering
    \begin{subfigure}[b]{0.32\linewidth}
    \centering
    \includegraphics[width=1.0\textwidth]{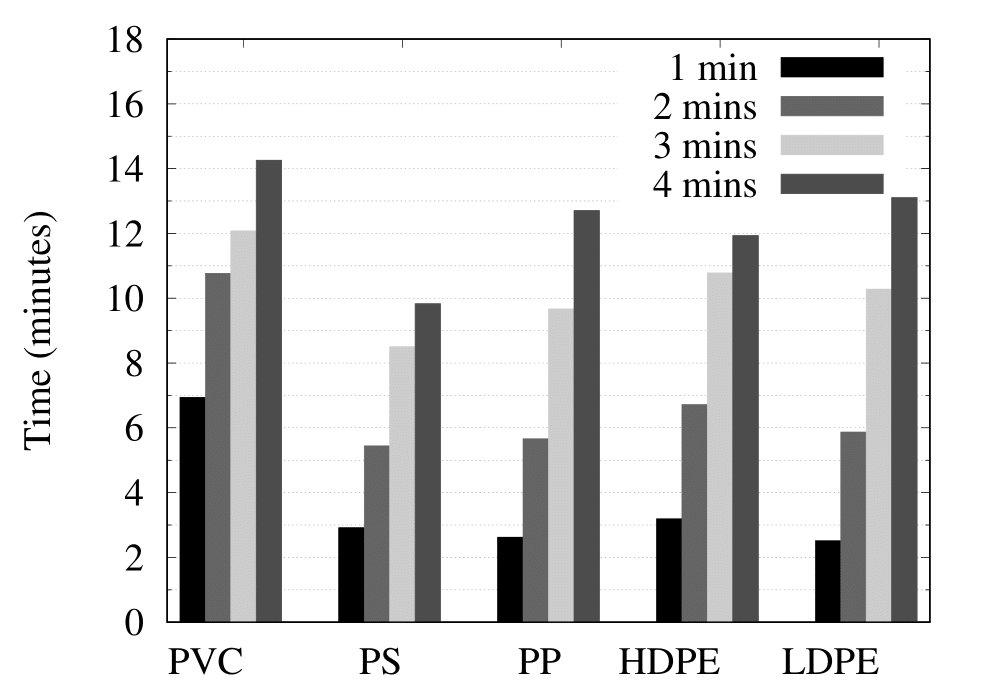}
    \caption{}\label{fig:plastics}
    \end{subfigure}
    \begin{subfigure}[b]{0.32\linewidth}
    \centering
    \includegraphics[width=1.0\textwidth]{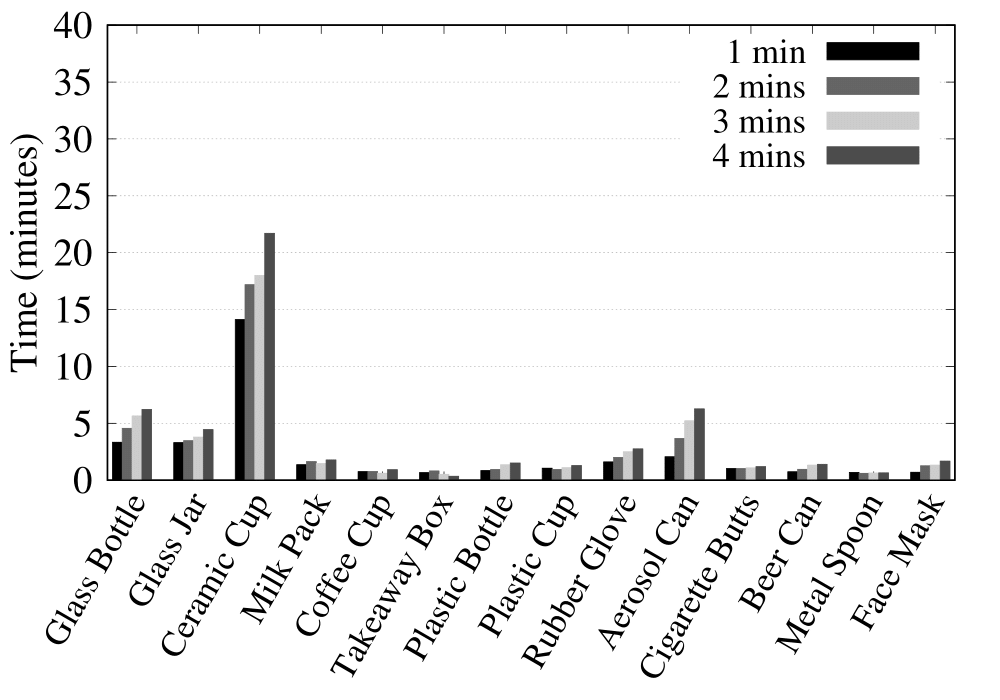}
  \caption{}\label{fig:allmaterials-thermometer}
 \end{subfigure}  
 \begin{subfigure}[b]{0.32\linewidth}
    \centering
    \includegraphics[width=1.0\textwidth]{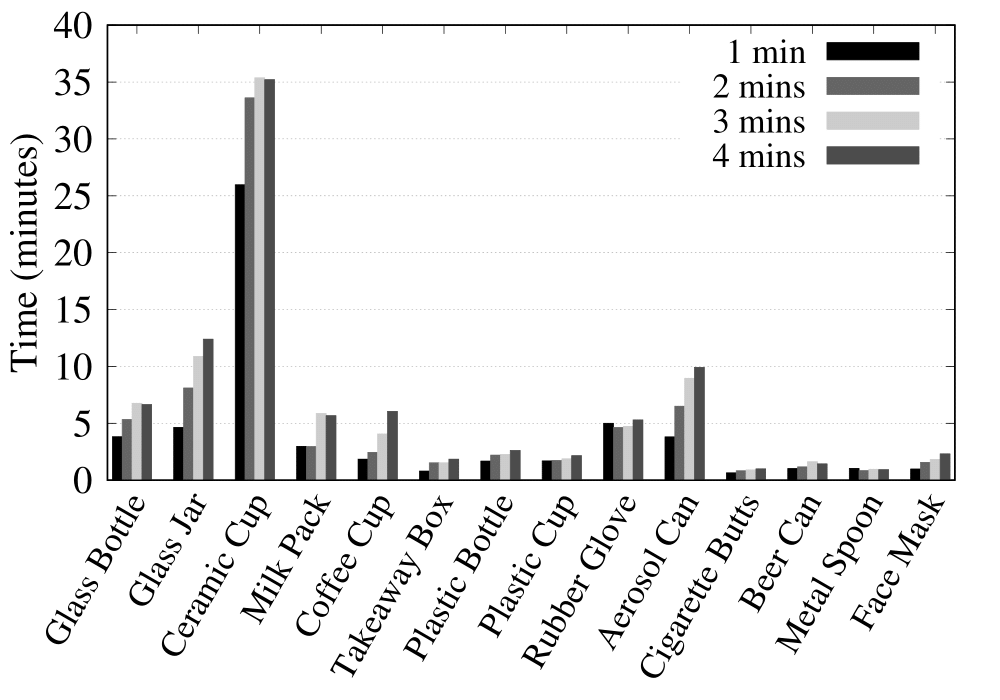}
    \caption{}\label{fig:allmaterials-cat60}
 \end{subfigure} 
 \caption{Dissipation time of thermal fingerprints in different plastic materials and waste material using two different devices: (a) Grouped by RIC code, (b) Thermometer scanner FLIR TG267 (baseline), and (c) Smartphone CAT s60.}
 \label{fig:allmaterials}
 \vspace{-.4cm}
\end{figure*}

\cparagraph{Results} The results in Figure~\ref{fig:allmaterials} are in line with the results for the plastic objects and indicate that the dissipation times differ across the objects and materials. Friedman test using object materials as experimental condition showed the differences of the materials to be statistically significant ($\chi^2(2)=48.83$, $p<.05$, $W=0.93$), demonstrating that thermal radiation can indeed characterize different object materials.

\section{\SystemName{} Pipeline}
\label{sec:methodology}

Aforementioned results demonstrated that dissipation of thermal fingerprints provides information that can be used to characterize different objects and identify their materials. We next briefly describe the sensing pipeline used as a solution to characterize everyday objects. \SystemName{} takes a sequence of thermal images taken from the object's surface as input and returns an estimate of the most likely material of that object. Next we describe the thermal image model which is used to classify object materials based on the dissipation time of thermal fingerprints.

\begin{figure}
	\centering
		\includegraphics[width=\columnwidth]{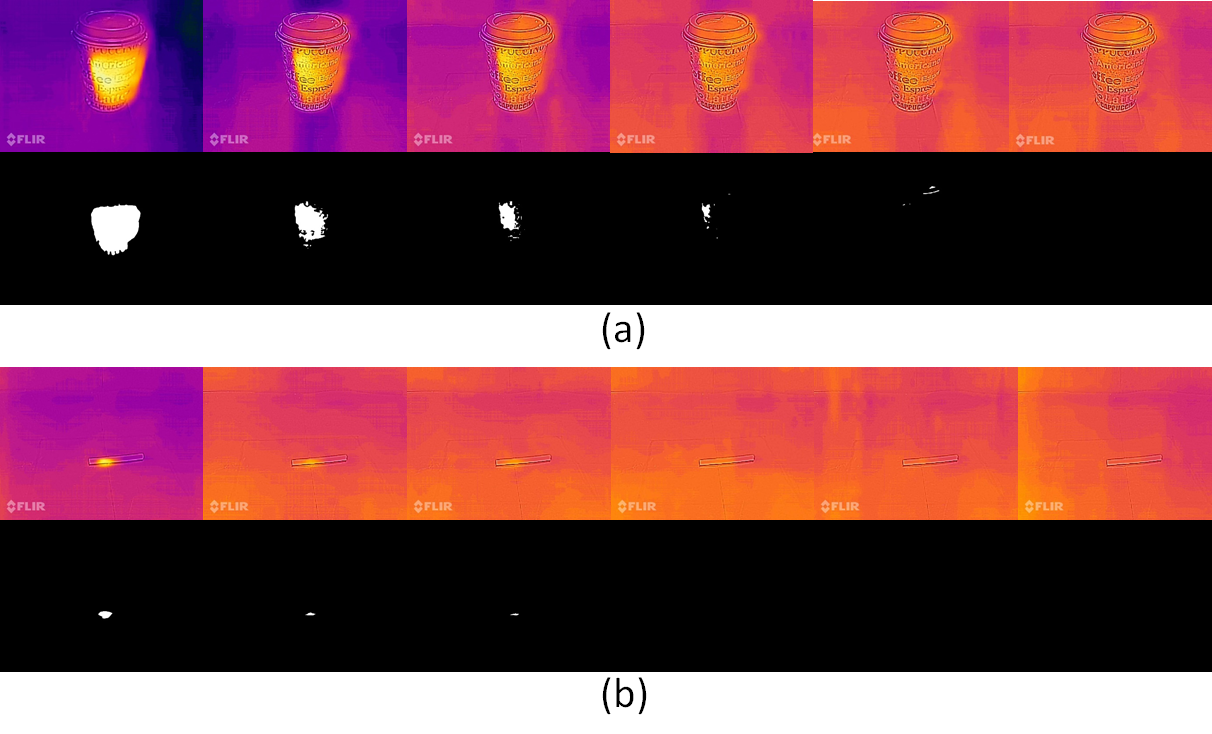}
	\caption{Dissipation time of thermal fingerprint for two different objects: (a) Cardboard cup and (b) Cigarette butt.}
	\label{fig:fading}
\end{figure}

\cparagraph{Preprocessing and Normalization} Since COTS thermal cameras are without cooling, they suffer from inaccuracies resulting from the camera overheating~\cite{malmivirta2019hot}. Other factors that influence measurement quality include misalignment between thermal and RGB pictures, internal recalibration of the camera, and low resolution. To mitigate these effects, we preprocess the thermal camera data by examining the background of consecutive images and removing images with significant dissimilarities. We also apply denoising on the images, normalizing the thermal values to a consistent scale (between $0$ and $255$). Such allows us to manipulate the images in gray scale. Figure~\ref{fig:fading} depicts the result of normalization procedure. The figures shows the sequence of thermal pictures over time (upper) and the corresponding image on the normalized scale (below) for two objects: (a) a cardboard cup and (b) a cigarette butt. This allows to isolate the thermal fingerprints from each object.

\cparagraph{Dissipation Rate} We estimate the dissipation rate of the thermal fingerprint from the normalized sequence of pictures as the function of area reduction of the thermal fingerprint given by the equation: 
\begin{equation}
RA = (A_{i} - A_{t})/A_{i},
\label{eq:reduction-area}
\end{equation}
\noindent where $RA$ is the reduction area percentage, $A_{i}$ is the initial area, and $A_{t}$ is the reduced target area~\cite{national2003fundamentals} (see Figure~\ref{fig:fading}). The reduction area between consecutive images is used to create vectors that model the dissipation time of thermal fingerprints for each object. \rev{To facilitate the training of machine learning classifiers, we consider fixed sized vectors. Note that once the fingerprint has dissipated, the target area is zero and thus this effectively corresponds to padding the vector with zeroes.}

\cparagraph{Implementation} Vectors with dissipation time are used as feature vectors and the type of object as label class. We then construct classification models using common machine learning techniques: Random Forests (RF), Support Vector Machines (SVM), and Multi-Layer Perceptron Classifier (MLPC).

\section{Robustness of Thermal Dissipation Fingerprints } 
\label{sec:experimentalsetup}

The experiments shown in Sec.~\ref{sec:materialselection} demonstrated that the dissipation characteristics of thermal fingerprints vary across different materials. Human body temperature varies across individuals and even within the same individual at different times of day~\cite{obermeyer2017individual}, which results in variations in the initial thermal fingerprints. Ensuring \SystemName{} can operate robustly against these variations is essential for ensuring the usefulness of \SystemName{} in practical use. In this section, we describe experiments where $18$ different individuals touch everyday objects, and we use the resulting thermal fingerprints and their dissipation to characterize the materials of the objects. The measurement setup is described in Section~\ref{ssec:testbed}. \rev{The experiments rely on measurements captured in an experimental testbed designed to capture video footage of thermal fingerprints for different objects materials. For this purpose, we used an off-the-shelf smartphone thermal camera and a thermometer scanner to measure the reference baseline. We consider three different materials which were chosen to have different shapes and sizes, but also to have sufficiently fast dissipation time to reduce the overall length of the study.}

\subsection{Experimental Setup}

\cparagraph{Experiment Design} 

We conduct a $3\times3$ within-subject design with holding pattern type and object type as independent variables. Both variables have three levels: Fixed-hold (FH), Natural-hold (NH) and Quick-hold (QH) for the former and Plastic bottle (BOTTLE), Cardboard cup (CUP) and Cigarette butt (CIGAR) for the latter. To eliminate order effects, whilst keeping the number of combinations manageable, holding pattern type was fully counterbalanced, whereas object type was counterbalanced following a Latin Square design, resulting in 
nine experimental conditions: 
(1) BOTTLE-FH, (2) CUP-FH, (3) CIGAR-FH; (4) BOTTLE-NH, (5) CUP-NH, (6) CIGAR-NH; (7) BOTTLE-QH, (8) CUP-QH, and (9) CIGAR-QH.

\begin{figure}[!t]
	\centering	\includegraphics[width=0.8\columnwidth]{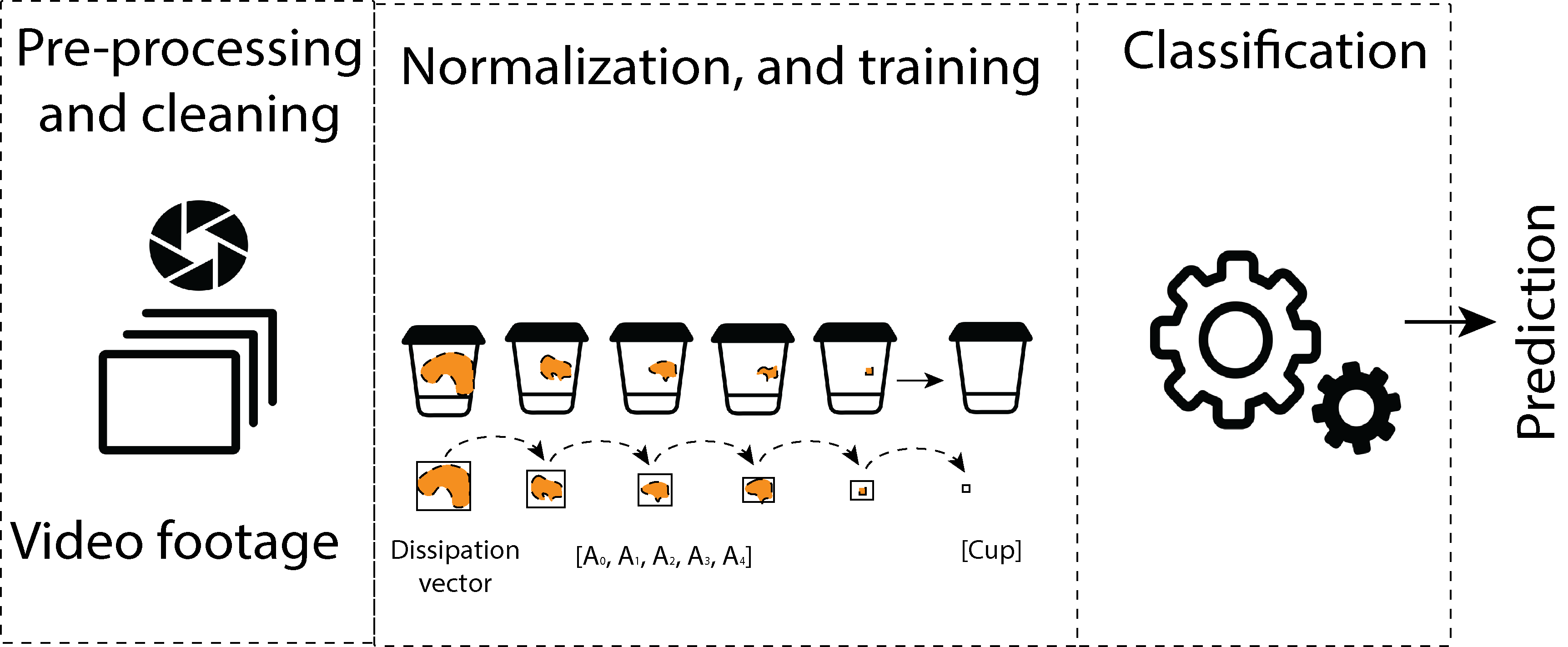}
	\caption{Processing pipeline of material classification based on dissipation time of thermal fingerprints.}
	\label{fig:approach}
	   \vspace{-.4cm}
\end{figure}

\begin{figure}
	\centering
		\includegraphics[width=0.65\columnwidth]{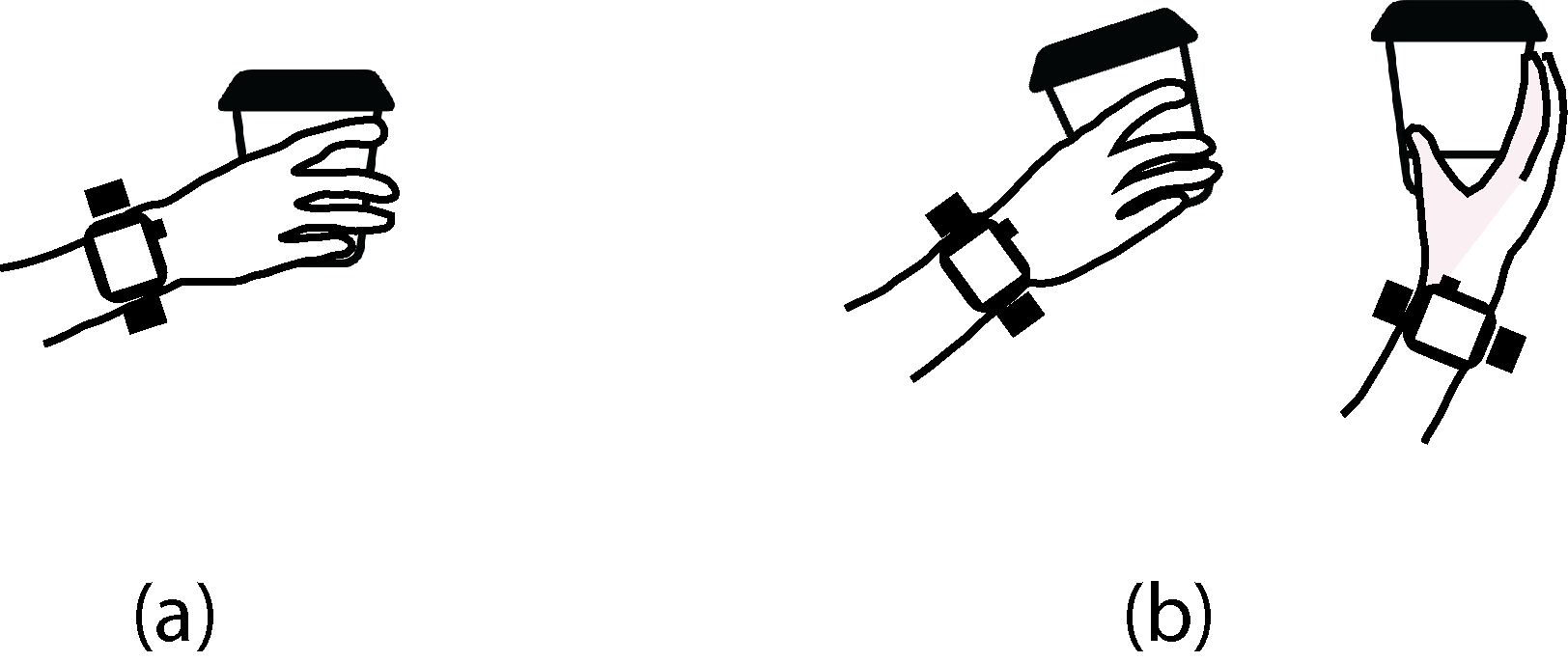}
	\caption{Experimental conditions: (a) Rigid interaction, used in Fixed-hold and (b) Free interaction, used in Natural and Quick holds.}
	\label{fig:conditions}
\end{figure}

In the~\textit{Fixed-hold} condition, objects were grabbed and held from a specific static position for one minute (Figure~\ref{fig:conditions}a). In the~\textit{Natural-hold} condition, objects were held freely for one minute, simulating usual everyday interactions with the object (Figure~\ref{fig:conditions}b). For instance, a participant holds the empty bottle for one minute while looking for a trash bin. In the~\textit{Quick-hold} condition, objects were held by participants freely for a \SI{10}{s} span.


\cparagraph{Participants} We recruited a total of N=$18$ participants (Males=$9$, Females=$9$) for the user study. Participants were students, admin staff and professionals from different fields, and nationalities, with little or no knowledge about thermal imaging. Their average age was $28\pm7.8$ years.

\cparagraph{Task} 

Participants were asked to hold objects and to simulate normal interactions with them. To produce data for natural interaction, we also asked the participants to contextualize a normal interaction context. When interacting with the BOTTLE, participants were asked to simulate drinking from the bottle and then looking for a trash bin to dispose an empty bottle. Similarly, participants were asked to stand while engaging in a short conversation with an acquaintance/friend for the CUP. Finally, for the CIGAR condition, participants were asked to simulate taking a cigarette from a cigarette box and then holding the cigarette from the filter while asking for a light. The cigarette was not lighted during the experiment.

\begin{figure}[!t]
	\centering
		\includegraphics[width=0.65\columnwidth]{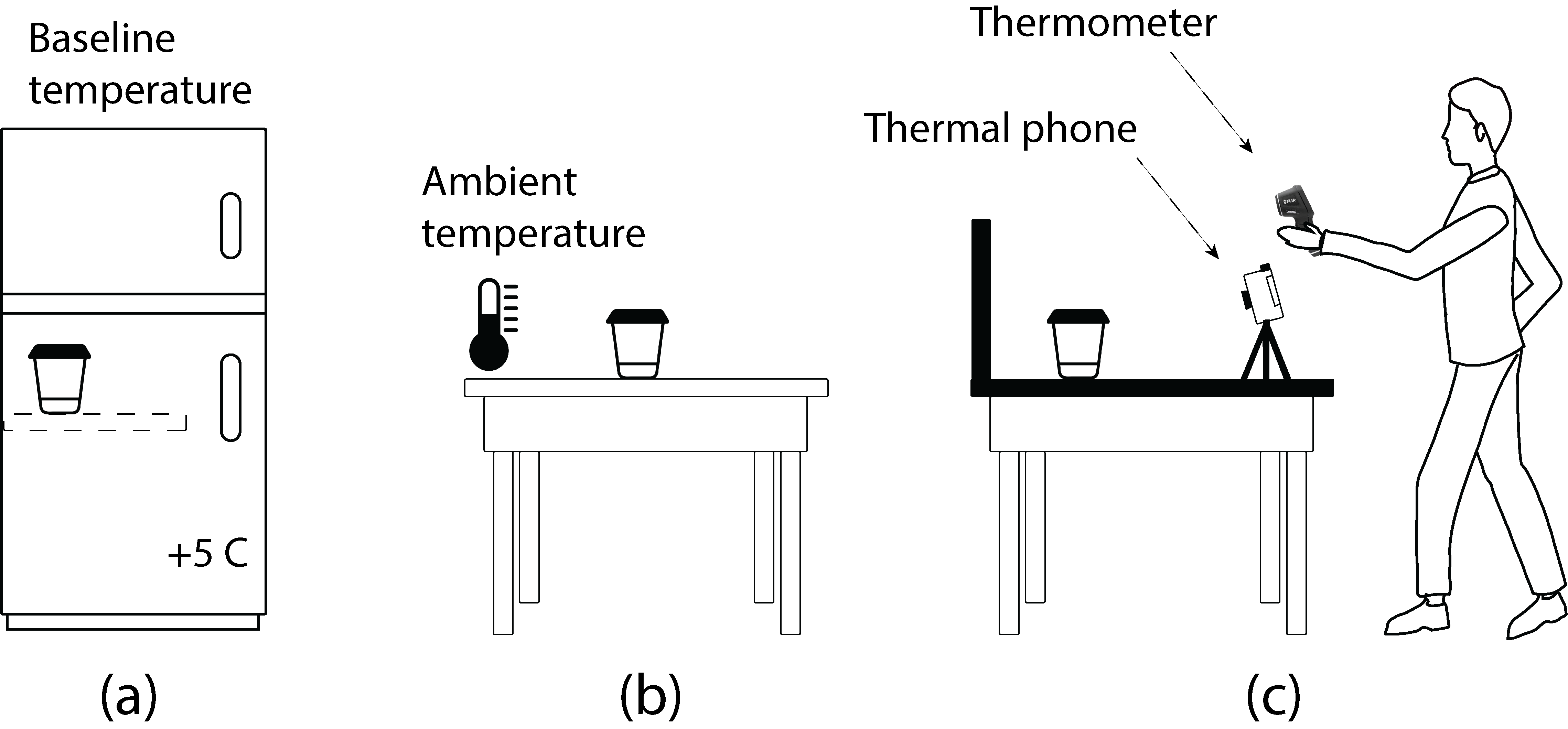}
	\caption{Experimental testbed and protocol steps: (a) Object obtains baseline temperature, (b) Object habituates to ambient temperature, and (c) Participant performs the experiment and puts the object in the marked target to measure its thermal fingerprint.}
	\label{fig:testbed}
\end{figure}

\cparagraph{Procedure} Before starting the experiment, each participant was invited to relax on a comfortable chair for $10$ minutes to enable the body temperature to acclimatize to the room's ambient temperature, which oscillated around \SIrange{22}{23.5}{\degreeCelsius} 
throughout the experiments. During this period, participants received a brief explanation of the study and signed an informed consent form, following local IRB regulations. Once the participant was ready to start the experiment, his/her/their body temperature was measured from the forehead using a clinically certified contactless optical thermometer (DR CHECK FC500). The nine experimental conditions were presented to participants. 
In each condition, the object was first placed inside an empty fridge with a temperature of \SI{5}{\degreeCelsius} for one minute (Figure~\ref{fig:testbed}a). This procedure rules out residual thermal radiation in the material between experiments and provides the material with a baseline temperature to make our results comparable across the participants. 

Kitchen tongs were used to take the object from the fridge to avoid heat transfer from humans to objects.
Next, the object was placed on a table for one minute to adapt it to the ambient temperature (Figure~\ref{fig:testbed}b). After that, participants carried out the corresponding experimental condition. Once finished, participants placed the object on a fixed marker drawn on a table with a black background and surface. The researcher conducting the study then used the CAT S60 to record video footage of the dissipation of the object's thermal fingerprint. In parallel, a thermometer scanner took thermal photos 
to serve as a reference baseline (Figure~\ref{fig:testbed}c). A black background helps to obtain clean video footage of thermal fingerprint without any thermal influence from objects in the surrounding environment. At the end of the experiment, we measured the participant's temperature from palm and finger to the objects using the thermal imaging scanner. The evaluation took place in one university room across two weeks in time slots between 11:00 and to 07:00 pm. 
Since human temperature varies during the day~\cite{obermeyer2017individual}, we considered only those times as they coincide with the working hours of the participants. For each participant, the overall experiment lasted  \SIrange{40}{45}{\minute}.

\subsection{Baselines}

As part of the experiments, we compare the recognition performance of \SystemName{} to two state-of-the-art techniques: deep learning based automated computer vision~\cite{white2020wastenet,ruiz2019automatic} and optical sensing~\cite{singh2016iot}. 

\cparagraph{Computer Vision} We train a state-of-the-art Convolutional Neural Network (CNN) model using the publicly available TrashNet dataset~\cite{aral2018classification}. We focus exclusively on the plastic materials category, which contains $626$ images of plastic objects for training the deep learning model. Plastics are malleable, so their accurate recognition is very sensitive to changes. The dataset has images where the individual pieces are shown against a white background. As such images do not match realistic recognition settings. To overcome this, we supplement the dataset with an additional $767$ images from the Japan Agency for Marine Earth Science and Technology (JAMSTEC) Deep-sea Debris Database dataset. We annotated the collected images manually by drawing a rectangle box around the object material in images. We labeled the TrashNet plastic items as "trash" and the JAMSTEC plastic items as "plastic". Both datasets were augmented by adding noise, hue, blue, horizontal flip, and vertical flip modifications to each original image, resulting in a total of $6985$ images for model training input. We created and trained the PlasticNet model using Google Collab server GPU, performing 100k iterations with a batch size of 12, running TensorFlow Lite 1.15. We used~\textit{ssd\_mobilenetv2\_oidv4} for the base training model with default hyperparameters.


\cparagraph{Light Reflectivity} As our second baseline, we consider reflectivity analysis of materials~\cite{singh2016iot} using a photoresistor connected to the analog input pin of an Arduino MEGA ADK. The photoresistor captures light changes based on its resistance exposure to the light intensity of the reflected material. As a light source, we rely on a red laser diode (wavelength \SI{650}{\nano\metre}). The object was located \SI{2}{\centi\metre} away from the light source, depicting a practical usage of the sensor in transport belts and smart bins~\cite{white2020wastenet}. We took measurements with sensor for different materials (selection is described in Section~\ref{sec:materialselection}), for one minute from two different random places in the object.

\section{Results} \label{sec:results}


We next demonstrate that \SystemName{} can characterize different object materials accurately using the measurements from the controlled user evaluation described in the previous section. We consider robustness against variations in the way humans interact with objects and variations in individuals, the overall classification performance for different materials, \rev{and the robustness of the detection to operate through other objects.}

\subsection{Differences in Thermal Fingerprints}

We first examine differences in thermal transfer from humans and the ensuing dissipation in the different objects. We separately consider the impact of the object and the hold type on thermal transfer. 

\begin{figure*}[htb]
\centering
    \begin{subfigure}[b]{.3\linewidth}
    \centering
    \includegraphics[width=1.0\textwidth]{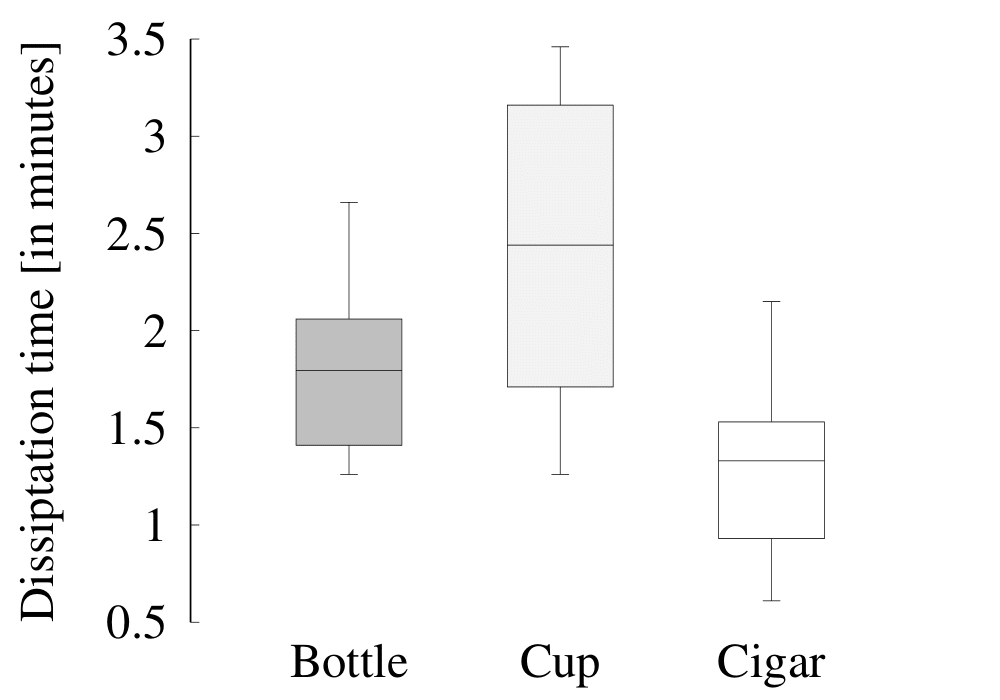}
   \caption{Fixed-hold}\label{fig:fixedhold}
  \end{subfigure}   
  \begin{subfigure}[b]{.3\linewidth}
    \centering
    \includegraphics[width=1.0\textwidth]{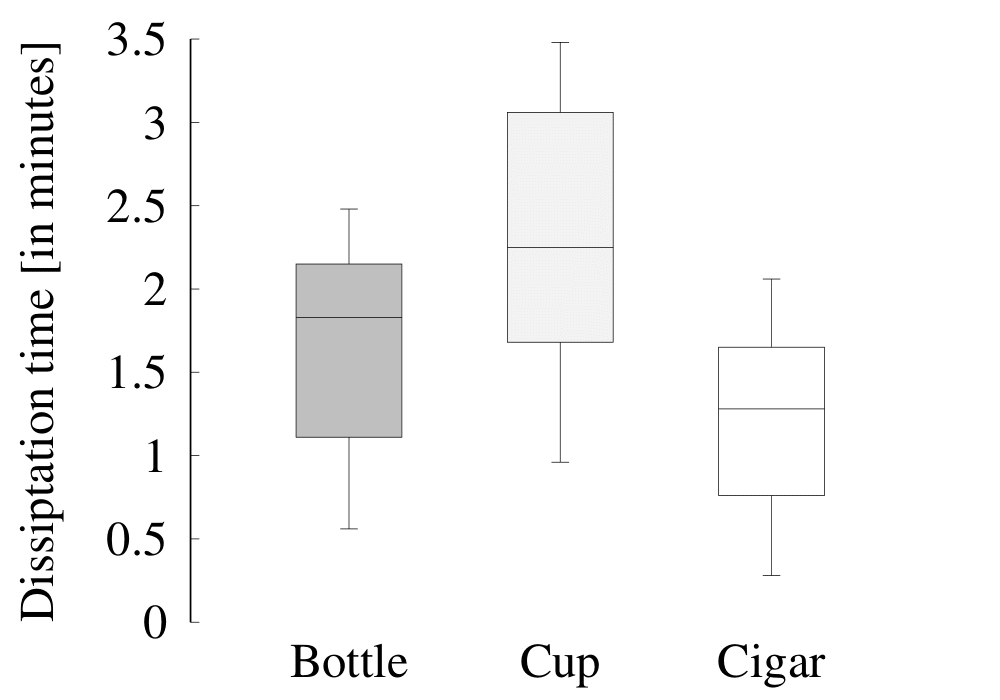}
   \caption{Natural-hold}\label{fig:naturalhold}
  \end{subfigure}   
  \begin{subfigure}[b]{.3\linewidth}
    \centering
    \includegraphics[width=1.0\textwidth]{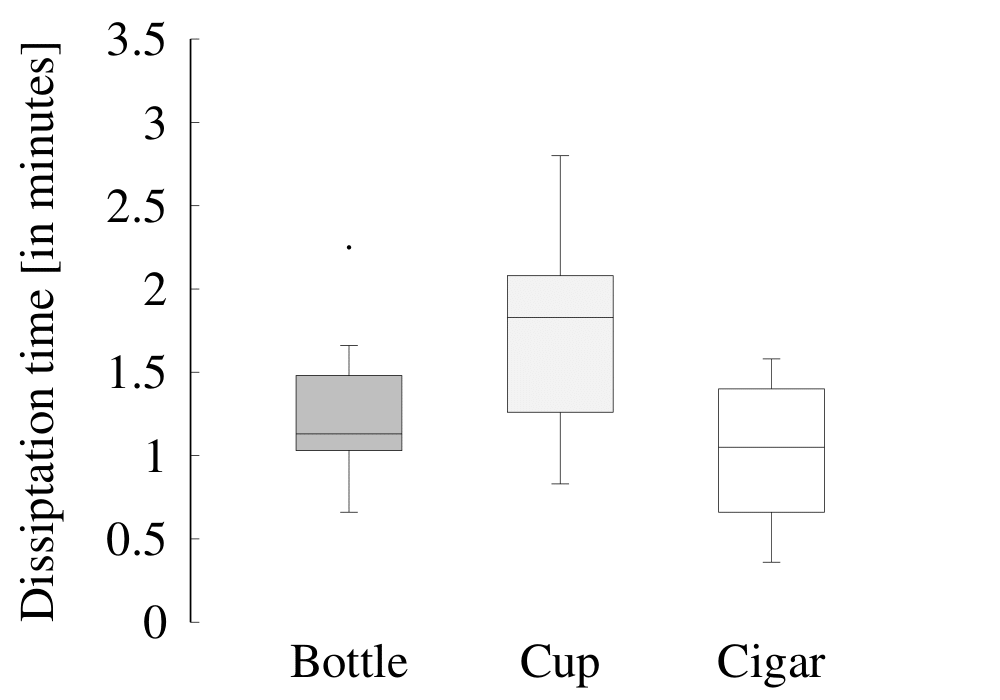}
    \caption{Quick-hold}\label{fig:quickhold}
  \end{subfigure} 
  \caption{Thermal transferred conditions applied over three different objects (plastic bottle, cardboard cup and cigarette butt).} 
  \label{fig:dissipation}
\end{figure*}

The dissipation times of the three objects under the different hold type conditions are shown in Figure~\ref{fig:dissipation}. From the figure, we can observe the differences in objects to be consistently different regardless of the hold type. Friedman Test using dissipation time and objects as experimental conditions shows the differences in objects to be significant for all of the three hold-type conditions: Fixed-hold ($\chi^2(2)=20.33$, p~$<.05$, $W=0.56$), Natural-hold ($\chi^2(2)=30.33$, p~$<.05$, $W=0.84$) and Quick-hold ($\chi^2(2)=25.04$, p~$<.05$, $W=0.64$). Pairwise post-hoc comparisons using Wilcoxon test (with Bonferroni correction for multiple comparisons) confirmed that the differences in dissipation times are statistically significant for all object pairs across the three hold types.

We next assess how the hold type affects the thermal fingerprints. 
Friedman test using dissipation time of each object and the three experimental conditions shows hold type to result in significant differences 
for the plastic bottle ($\chi^2(2)=12.44$, p $<.05$, W=$0.34$) and the cardboard cup ($\chi^2(2)=16.48$, p $<.05$, W=$0.45$), but not for the cigarette butt. Post-hoc comparisons indicate the thermal fingerprints for the Quick-hold pattern to significantly differ from those in the Fixed- and Natural- hold conditions. Results imply that \textit{differences} in thermal fingerprints contain significant variation across objects regardless of how users touch or interact with them. Still, the dissipation times are impacted by the time the user holds the item. This is expected, as the time the user touches the object affects the extent of heat that can transfer and thus controls dissipation speed.

We also separately analyzed whether differences in body temperature across different body parts can affect the thermal fingerprint by comparing the thermometer results across the three measurement locations (forehead, hand palm, finger tips). Friedman test using part of body as experimental conditions showed significant differences for temperature ($\chi^2(2)=29.66$, $p<.05$, $W=0.82$). Post-hoc comparisons (Dunn-Bonferroni) verified that the differences were statistically significant ($p<.01$) between forehead and hand-palm and between forehead and finger-tips. The average temperature for the different parts of body were: forehead \SI{36.33}{\degreeCelsius}, hand-palm \SI{30.16}{\degreeCelsius}~and finger-tips \SI{30.87}{\degreeCelsius}. 
These results show that fingers and hand palm generally induce similar heat transfer, further validating the robustness of the thermal fingerprints against the way people interact with the objects. The difference to forehead temperature, in turn, suggests that the hand palm and fingertips react to ambient temperature instead of correlating with the internal body temperature -- as is the case for the forehead measurements. We note that the higher sensitivity of finger tips and hand palm also means that it can be potentially used to extract insights about human activity. For example, interactions with a smartphone can result in heat caused by battery temperature to transfer into the hand, and the contents of a hot drink can similarly affect the temperature.  




Overall, the results indicate that human touch transfers a sufficient amount of heat, making it possible to characterize objects based on touch without resorting to specialized technology. However, the results show that the dissipation times are affected by the time the user interacts with the objects -- and other factors as will be shown in the next subsection -- implying that relative differences in dissipation characteristics should be used instead of the exact dissipation times for characterizing materials.

\subsection{Other Factors Influencing Thermal Dissipation}


Thus far we have shown that the characterization of object materials with thermal radiation depends on exposure time, but not on the location where the object is being touched at. In this section, we investigate other factors that influence the transfer of thermal radiation from the human body to objects.



\cparagraph{Gender and Temperature} Body temperature generally influences thermal radiation transferred to objects. We next analyze whether the gender of the participants affects thermal transfer, i.e., whether there are significant differences in thermal transfer between female and male participants. We separately assessed effects for the hand palm and finger tips. Kruskal-Wallis tests, using gender and part of the body as experimental conditions, show that there are significant differences in thermal transfer only for finger tips ($\chi^2(2)=5.08$, p $<.05$). This is most likely the result of differences in the size of the contact area with males typically having larger fingertip size~\cite{acree1999there}.


When using gender and objects as experimental conditions, Kruskal-Wallis tests show there to be significant differences in the dissipation time for all three objects: cigarette butt ($\chi^2(2)=3.94$, p $<.05$), plastic bottle ($\chi^2(2)=12.17$, p $<.05$) and cardboard cup ($\chi^2(2)=7.75$, p $<.05$).
These results indicate that the thermal fingerprint's dissipation time depends on temperature and that it is possible to identify whether a female or male individual has touched the object. While this result does not change the fact that objects can be characterized with thermal radiation, \rev{it is also important to highlight the potential privacy implications. Indeed, the results show that thermal radiation can disclose additional information about the humans interacting with the objects.} For example, it could be possible to use thermal fingerprints to compare household waste sorting practices between genders.  



\begin{figure}[!t]
\centering
    \begin{subfigure}[b]{.45\linewidth}
    \centering
    \includegraphics[width=1.0\textwidth]{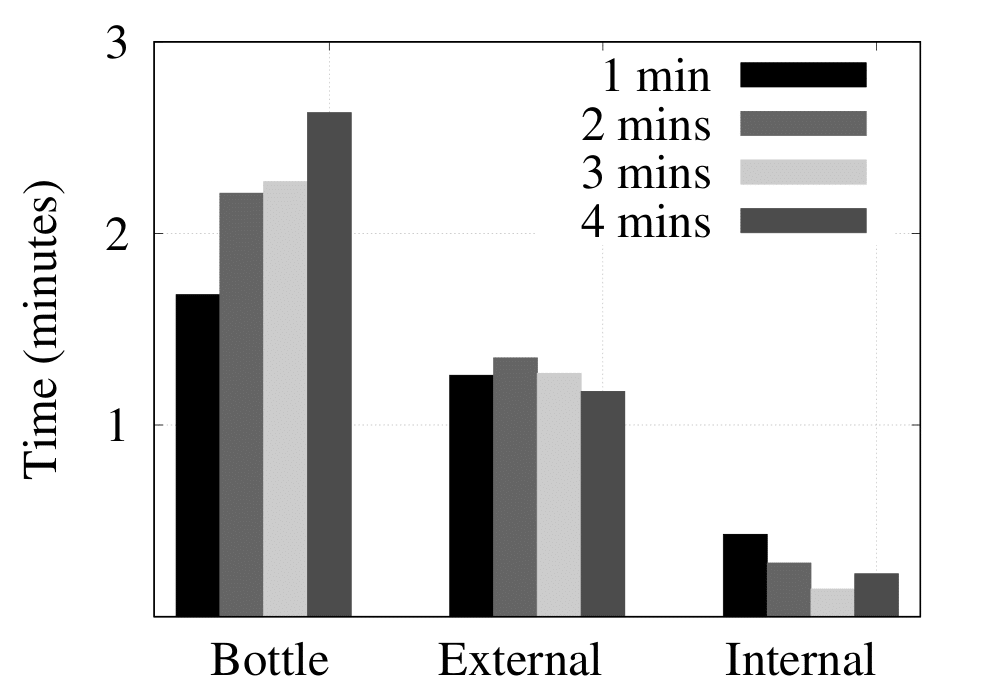}
   \caption{CAT s60}\label{fig:bottle-cats60}
  \end{subfigure}   
  \begin{subfigure}[b]{.45\linewidth}
    \centering
    \includegraphics[width=1.0\textwidth]{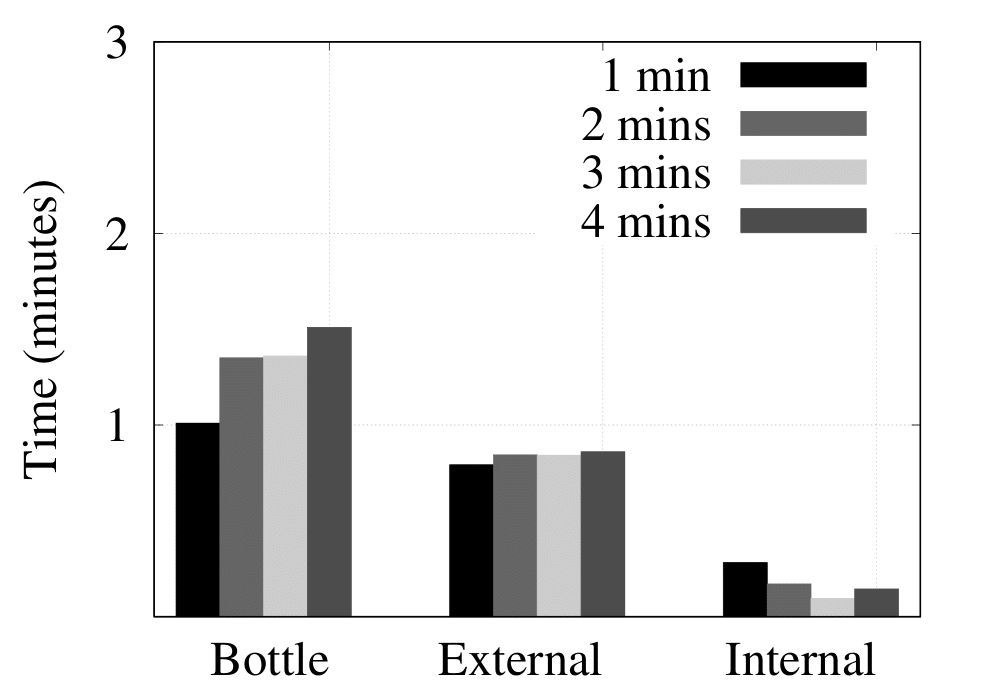}
    \caption{Thermometer Scanner}\label{fig:bottle-thermometer}
  \end{subfigure} 
  \caption{Influence of internal and external temperatures.} 
  \label{fig:factors}
	   \vspace{-.4cm}
\end{figure}

\cparagraph{External Temperature of Ambient Environment} Surrounding temperature of the object directly influences the dissipation time of the thermal fingerprint. We quantity the influence of this factor through additional small scale experiments. First, the BOTTLE was held by a human hand for different periods at an ambient temperature of \SIrange{22}{23.5}{\degreeCelsius}. Next, the BOTTLE was placed inside a colder environment (fridge with a temperature of \SI{5}{\degreeCelsius}). We then measured the dissipation time of the thermal fingerprint when changing from ambient to colder environment using both the CAT S60 and the thermometer scanner. Figure~\ref{fig:factors} shows the results. We also include the thermal fingerprint's dissipation time in the ambient environment for comparison purposes (baseline). We notice the total dissipation time of the thermal fingerprint is halved when changing to a colder environment. Still, overall the differences in change patterns remain consistent for the different objects. This suggests that the environment affects the fingerprints. Note that the magnitude of this change is proportional to heat difference and impacts all objects equally. Hence, incorporating the ambient temperature in the thermal dissipation fingerprints is sufficient to overcome potential issues from differing temperatures.  

\cparagraph{Internal Temperature absorbed from Contents} Besides the ambient temperature of the surrounding environment, objects can also be influenced by the thermal radiation resulting from the contents of the object. For example, a cardboard cup can be filled with a hot or a cold drink. To investigate this further, we fill the BOTTLE with water with a temperature of \SIrange{21.2}{21.5}{\degreeCelsius}. Before the experiment, we place the BOTTLE inside a fridge (\SI{5}{\degreeCelsius}) to eliminate thermal radiation carryover effects between experiments. We then compare the dissipation times of an empty and filled bottle in ambient temperature (\SIrange{22}{23.5}{\degreeCelsius}). Figure~\ref{fig:factors} depicts the results. We observe that the internal radiation impacts the thermal fingerprint of the BOTTLE. This is relevant to identify end-products that have not been fully consumed. In practice, such cases should be modeled as a separate (mixed) object to ensure the model can distinguish the pure material from those cases where there are no contents inside the object. Compared with the thermal fingerprint of the empty BOTTLE, we can observe that the thermal fingerprint of the filled water BOTTLE dissipates faster as a result of the larger difference with the environment. 


\cparagraph{Distance between Object and Thermal Camera} In the experiments thus far, the distance between the thermal camera and the objects has been fixed at \SIrange{30}{35}{\centi\metre} (baseline). We next analyze the effect of longer distances by considering three additional distances: \SI{70}{\centi\metre} (distance-1), \SI{105}{\centi\metre} (distance-2) and \SI{210}{\centi\metre} (distance-3). 
We focus exclusively on the BOTTLE with a Fixed-hold setup of one minute. At \SI{70}{\centi\metre} distance, the dissipation time does not change significantly (average time = $1.13$ min) for the CAT S60. At longer distances, we observed higher variations in the dissipation time with the CAT S60. At \SI{105}{\centi\meter} distance the average time was $0.76$ minutes and at \SI{210}{\centi\metre} the average time was $0.26$ minutes. The resolution of the thermal camera in the CAT S60 is $80\times60$, which seems to be sufficient for up to a meter. The thermometer scanner has a higher resolution $160 \times 120$, but a slower frame rate (\SI{6.67}{\hertz} vs. \SI{8}{\hertz}). For a \SI{210}{\centi\meter} distance, the thermometer fails to observe a proper thermal fingerprint. A higher resolution is thus not guaranteed to extend the operational range of \SystemName{} as also the frame rate needs to be considered.

\cparagraph{Temperature Sensitivity} We next analyze the sensitivity of the dissipation times to slight variations in temperature. 
To accomplish this, we further analyze the BOTTLE material in the fixed-hold condition, as above. We use a JANOEL18S incubator with adjustable temperature to achieve controlled changes in the temperature. We put the object inside the incubator at a constant temperature for $15$ min to ensure that the entire object has exactly the same temperature. We expose the material to temperatures ranging from \SIrange{36}{39}{\degreeCelsius}, corresponding to normal and elevated temperature levels in a human. After the object was heated up, it was transferred to the testbed that was used to record footage. We then proceed to measure the dissipation time of the thermal fingerprint. The room temperature that the object had to acclimatize ranges from \SIrange{23}{23.5}{\degreeCelsius}. The dissipation times for the different temperatures were: 3.33 min for \SI{36}{\degreeCelsius}, 3.73 min for \SI{37}{\degreeCelsius}, 4.23 min for \SI{38}{\degreeCelsius}, and 4.34 min for \SI{39}{\degreeCelsius} respectively. The dissipation time is expected to be a function of the temperature difference between the object and the environment, and our results confirm that these subtle temperature differences can be captured robustly using a commercial-off-the-shelf thermal camera. 
While accurate human temperature is not possible to estimate, we envision that our approach can be used to detect abnormalities in human temperatures. For instance, instead of monitoring the face of people with thermal cameras at an airport, it could be possible to detect abnormal temperatures by monitoring the tangible objects that people touch while passing a security check. 

\subsection{Dissipation Time Classification Performance}
\label{ssec:classification}

We next demonstrate that our approach can support the coarse-grained classification of object materials based on the dissipation time of thermal fingerprints -- and other contextual factors. As described in Section~\ref{sec:methodology}, we considered three classification techniques: Random Forest (RF), Support Vector Machine (SVM), and Multi-Layer Perceptron (MLP). The results of the classification experiments are shown in Table~\ref{tab:classification}. When only the thermal fingerprint is available, the highest classification accuracy for material detection is approximately  $83$\%. 
Incorporating information about the hold pattern type does not improve the results, suggesting that the way people grasp the materials does not impact performance. 
In contrast, when having information about whether the person is male or female, we can observe that the accuracy to predict materials improves up to $86$\%. Similarly, when attempting to predict whether the user is male/female, we can observe that dissipation time and material information provide a high accuracy estimation of around $78$\%. 



\begin{table}
\centering
\caption{Material classification accuracy (\%) in different experimental conditions. Model data $\rightarrow$ Predicted. Classification Method: Random Forest (RF), Support Vector Machine (SVM) and Multi-layer Perceptron (MLPC).}
\label{tab:classification}
\resizebox{0.8\textwidth}{!}{%
\begin{tabular}{lcccc}
\toprule
\textbf{Test} & \textbf{RF} & \textbf{SVM} & \textbf{MLPC} & \textbf{Average} \\
\hline 
\textbf{Predicting Material (M)}
 &  &  &  \\
(Vector) $\rightarrow$M&90.9&77.3&81.8&\textbf{83.3}\\
(Vector, Context)$\rightarrow$M&90.9&77.3&81.8&\textbf{83.3}\\
(Vector, Gender)$\rightarrow$M&90.9&86.4&81.8&\textbf{86.4}\\
(Vector, Context, Gender)$\rightarrow$M&86.4&81.8&81.8&\textbf{83.3}\\
\quad\textit{Average}&\textit{89.8}&\textit{80.7}&\textit{81.8}&\textbf{\textit{84.1}}\\
\textbf{Predicting Context, Gender}
 &  &  &  \\
(Material, Vector)$\rightarrow$Context&77.3&81.8&72.7&\textbf{77.3}\\
(Material, Vector)$\rightarrow$Gender&77.3&77.3&81.8&\textbf{78.8}\\
\quad\textit{Average}&\textit{77.3}&\textit{79.6}&\textit{77.3}&\textbf{\textit{78.1}}\\

\bottomrule
\end{tabular}%
}
\end{table}

\subsection{Comparison Against Other Approaches}

\begin{figure}[!t]
\centering
    \begin{subfigure}[b]{.5\linewidth}
    \centering
    \includegraphics[width=0.8\textwidth]{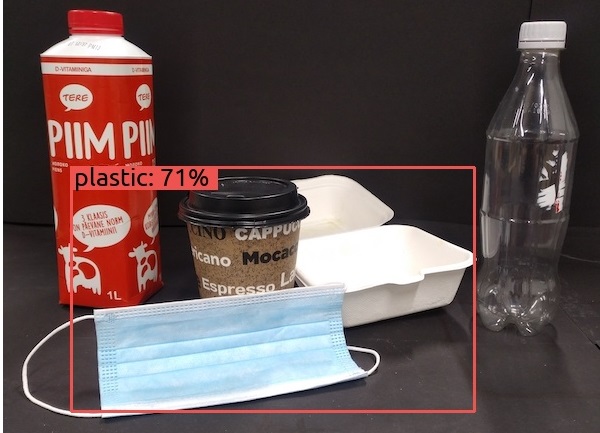}
   \caption{}\label{fig:baseline1}
  \end{subfigure}   
  \begin{subfigure}[b]{.45\linewidth}
    \centering
    \includegraphics[width=1.0\textwidth]{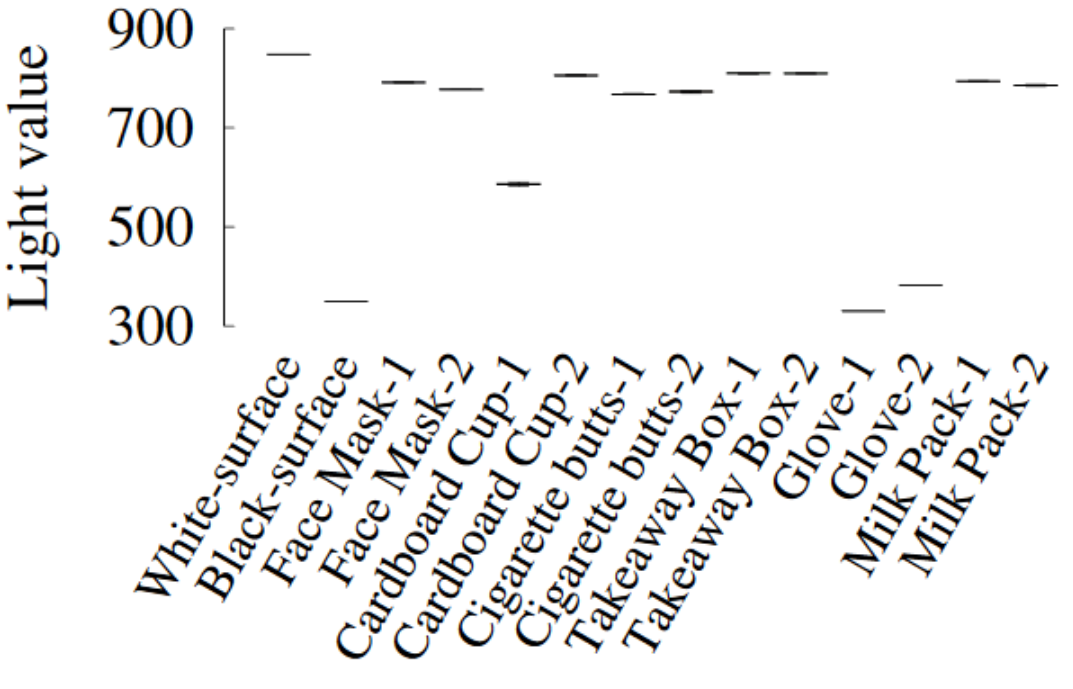}
    \caption{}\label{fig:baseline2}
  \end{subfigure} 
  \caption{Baselines: (a) Trained model PlasticNet to identify and separate plastics from other object materials, and (b) Light reflectivity values of different materials measured with a photo-resistor.} 
  \label{fig:baselines}
  \vspace{-.4cm}
\end{figure}

Next we compare \SystemName{} against the two baselines: computer vision and optical sensing. 
We test  computer vision using 31 images depicting real tossed plastic objects~\cite{prata2020covid}. In total, 33 separate plastic items were present in the 31 images. The deep learning model managed to identify 23 of the 33 items (69.7\% in accuracy). \rev{Note that the goal is not to directly compare our approach against computer vision as they effectively address different problems. Indeed, the vision-based approach does not extract any internal characteristics of the objects and hence the only way it can be used in material recognition is to map specific item types to materials (e.g., drink bottles are typically made of PET). This approach works reasonably when the objects are sufficiently distinctive but fails in everyday use cases where the shape of the objects has changed, e.g., tossed objects often have lost their original shape, and other visual characteristics may similarly have undergone significant changes. Computer vision is also prone to occlusions and cases where the objects are only partially visible. These issues are highlighted in Figure~\ref{fig:baselines}a.  In practice computer vision and thermal dissipation can complement each other, e.g., autonomous ground vehicles could use computer vision to detect litter objects from afar, navigate next to the object, and use thermal dissipation to determine the material of the object. }


Figure~\ref{fig:baselines}b shows the results for the second baseline, light reflectivity. 
The low variation in reflectivity values indicates that light can accurately characterize different materials. However, we also observe  that different parts of the same material can be characterized very differently, e.g., Cardboard Cup-1 and Cardboard Cup-2. This is because objects comprise of different materials and colors, which can affect reflectance. Another limitation of this approach is the need for the sensors to contact the material to classify it accurately.

\subsection{Effect of Thickness on Thermal Dissipation}
\label{ssec:thickness}

The results thus far have investigated the case where the camera can observe the objects that the user interacts with directly but in many practical scenarios the objects will be partially or fully occluded by other objects. \rev{We next demonstrate that thermal dissipation can be captured even in cases where the objects are covered by other objects. Naturally the thermal dissipation is affected by the materials that cover the object and hence the purpose of these experiments is to demonstrate the potential of our technique to support practical applications rather than to argue it can be used to detect materials through cover. }

\begin{figure*}
	\centering
		\includegraphics[width=1.0\textwidth]{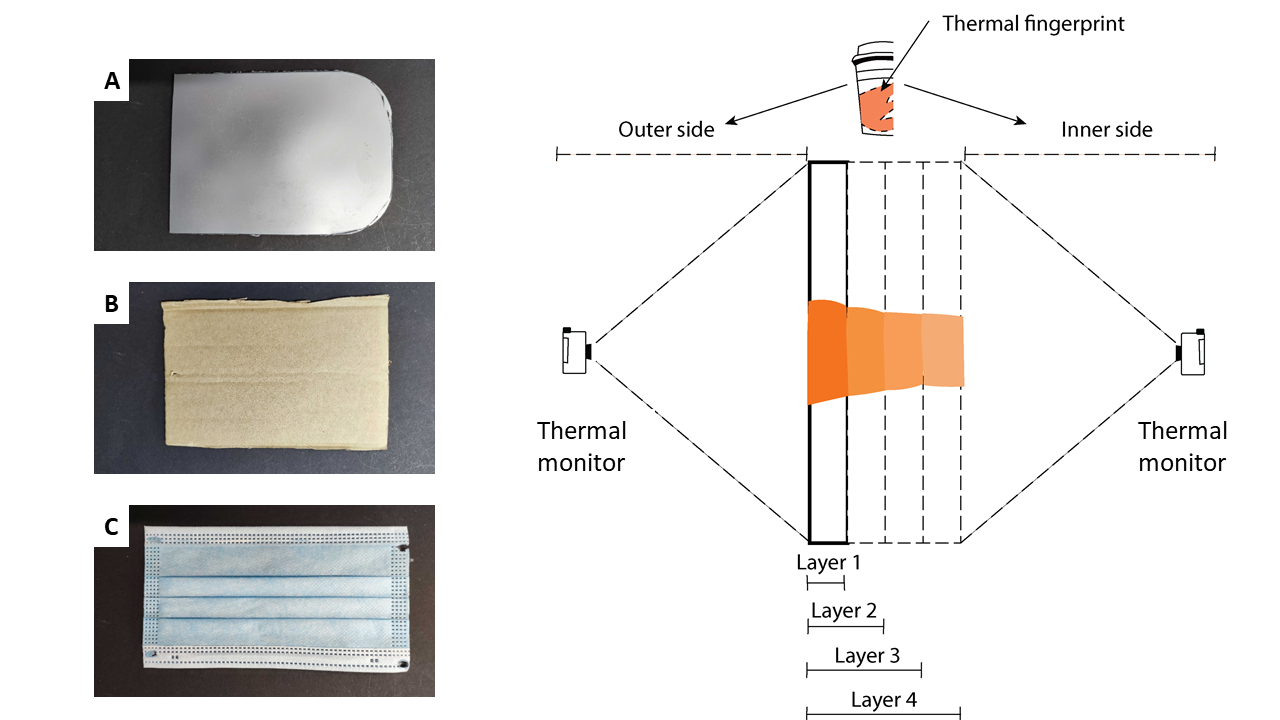}
	\caption{Thickness of object materials analyzed with thermal dissipation fingerprint.}
	\label{fig:thickness}
\end{figure*}

\cparagraph{Experimental setup} Thermal radiation is absorbed from the outer side of an object material to its inner side as it dissipates. We conduct an experiment where thermal radiation from the outside of an object is absorbed and dissipated to its interior and used to model the thermal radiation of the object. As material influences thermal dissipation, as we have demonstrated previously, we design specialized samples to analyze the effect of thickness for thermal dissipation. These samples are developed from everyday household objects. In this experiment, we selected a plastic container (Plastic), a cardboard box (Cardboard), and a face mask (Mask) as baseline thickness for our samples (Figure~\ref{fig:thickness} shows the samples). We then aggregated multiple samples of the same object in layers to obtain samples with different thicknesses. \rev{Note that the overall penetration of the thermal radiation depends on two factors: (i) the thickness of the material and (ii) the characteristics of the material. These two factors together determine the thermal resistance of a material~\cite{mahlia2007correlation} which controls how much heat passes through, and thus also how much thermal radiation can be captured on the other side of the covering material.  In the experiments, we} consider a maximum of four layers as we found that thermal dissipation reduces significantly after this stage. The maximum thickness in the experiments is around \SI{2}{mm} which corresponds to the thickness of many everyday consumer products. For example, a plastic water bottle typically is between \SIrange{1}{2}{mm} thick. The material thickness in each layer is shown in Table~\ref{tab:tablethickness}. We rely on a Fujisan FJS025 electronic micrometer screw gauge having an accuracy of \SI{0.001}{mm} and range from \SIrange{0}{25}{mm} to measure the thickness of the objects.

\cparagraph{Procedure} We measure thermal dissipation fingerprint from touch on the topmost material for one minute. After this, thermal dissipation is measured from the inner side of the object. This experiment is repeated three times per sample and multiple layers. We also measure the thermal dissipation time of the outer side of the object as a reference. The average body temperature of the human subject holding the object fell within \SIrange{36.4}{36.7}{\degreeCelsius}, and the ambient temperature was from \SIrange{22}{24}{\degreeCelsius}. Experiments were conducted for six days between 12:00 am and 14:00 pm.

\begin{figure}[!t]
\centering
    \begin{subfigure}[b]{.45\linewidth}
    \centering
    \includegraphics[width=1.0\textwidth]{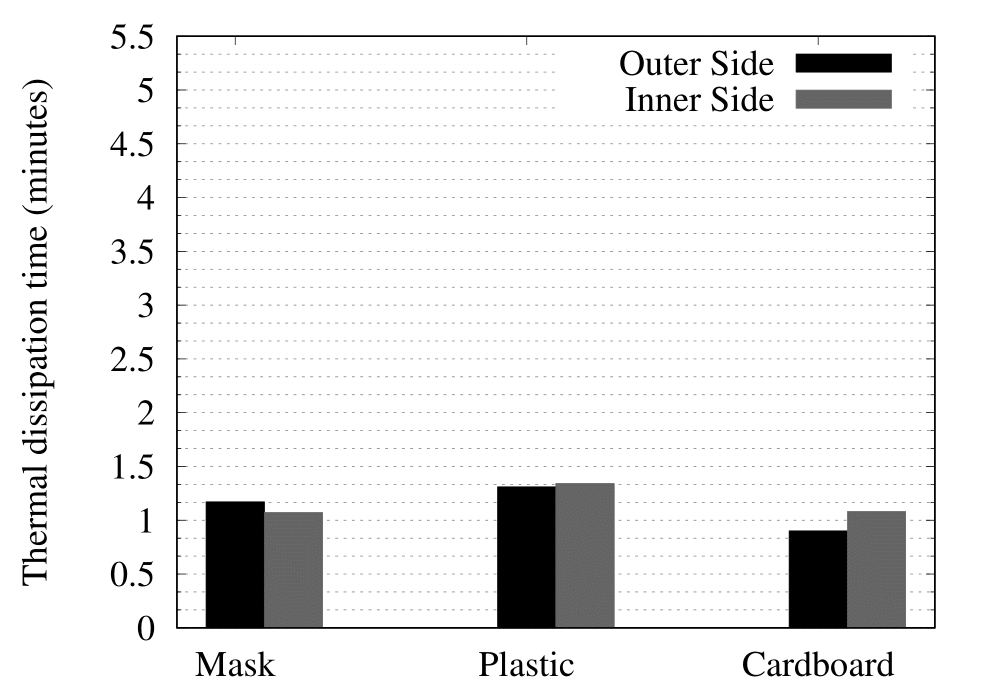}
   \caption{One Layer Thickness}\label{fig:thicknesslayer1}
  \end{subfigure}   
  \begin{subfigure}[b]{.45\linewidth}
    \centering
    \includegraphics[width=1.0\textwidth]{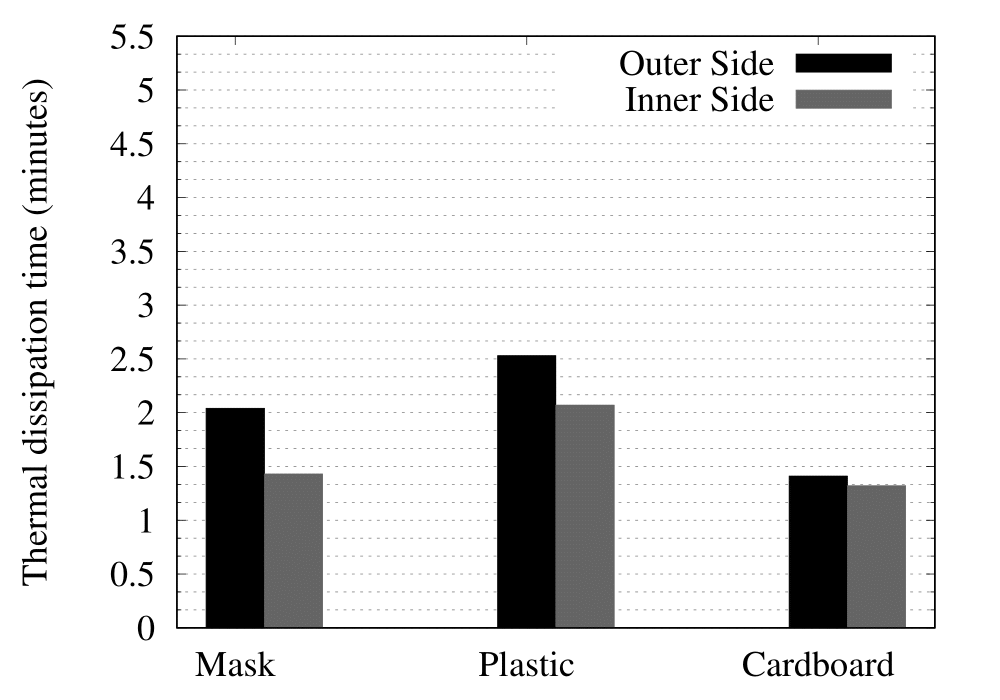}
    \caption{Two Layer Thickness}\label{fig:thicknesslayer2}
  \end{subfigure} 
  \begin{subfigure}[b]{.45\linewidth}
    \centering
    \includegraphics[width=1.0\textwidth]{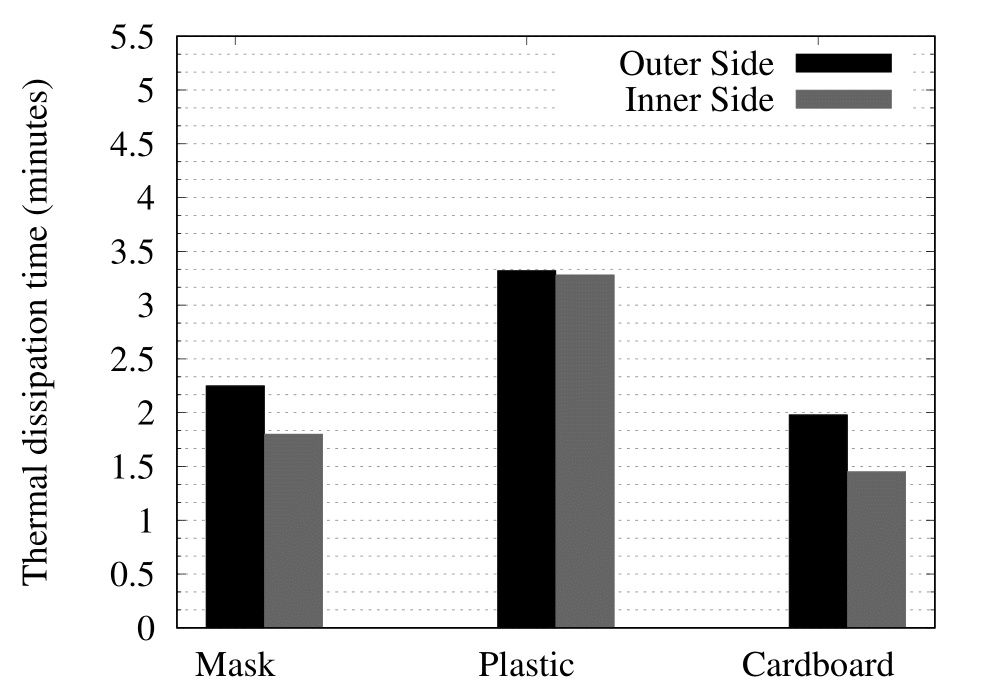}
    \caption{Three Layer Thickness}\label{fig:thicknesslayer3}
  \end{subfigure} 
  \begin{subfigure}[b]{.45\linewidth}
    \centering
    \includegraphics[width=1.0\textwidth]{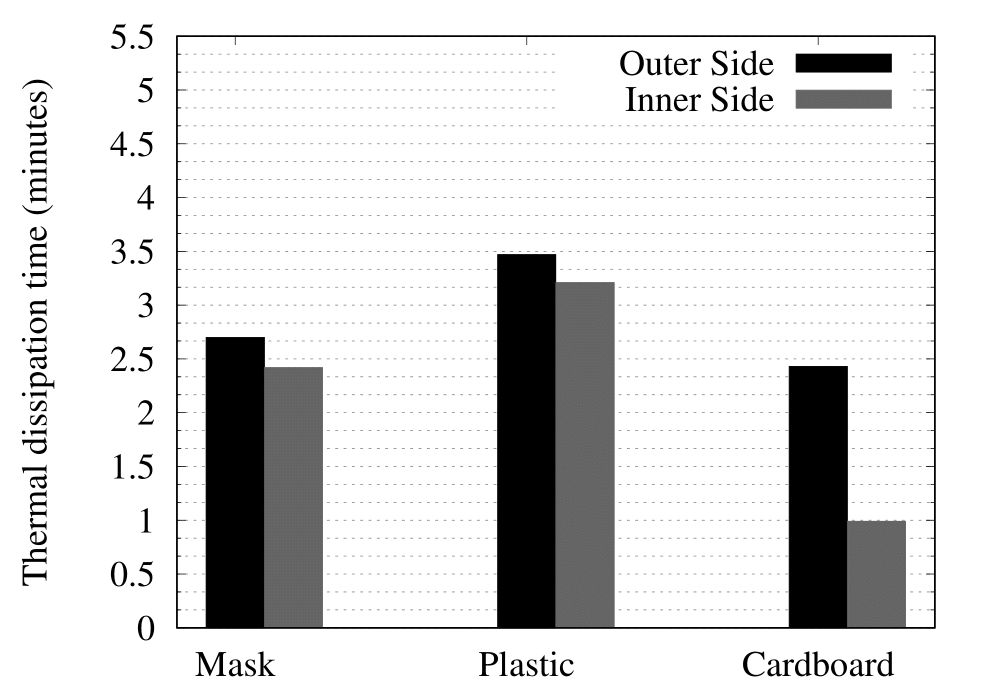}
    \caption{Four Layer Thickness}\label{fig:thicknesslayer4}
  \end{subfigure} 
  \caption{Effect on thermal dissipation with increase in the thickness of material.}
  \label{fig:thickness1}
	   \vspace{-.4cm}
\end{figure}

\cparagraph{Results} Figure~\ref{fig:thickness1} shows the thermal dissipation estimated from the inner side of the object. The figure also includes the thermal dissipation fingerprint of objects from their outer side. From the results, we can observe \rev{two effects. First, the dissipation time increases as a result of larger surface area. Second, the dissipation time from the inside is smaller than the dissipation time from the outside with the total difference depending on the type and thickness of the material (i.e., higher thickness results in higher difference). Indeed, the difference is smaller for the plastic bottle compared with the face mask or the cardboard. This highlights how the effect of thickness depends also on the material that is covering the object. Nevertheless,} the results show that it is possible to observe that thermal radiation from human touch is visible even in samples with multiple layers. \rev{Besides demonstrating the robustness of the dissipation fingerprints against partial or full occlusion}, the result also has potential to support integration of \SystemName{} into everyday objects; see Section~\ref{sec:Discussion}.



\begin{table}[]
\centering
\caption{Material thickness in each layer for Cardboard, Plastic and Mask respectively.}
\label{tab:tablethickness}
\resizebox{0.8\textwidth}{!}{%
\begin{tabular}{|l|l|l|}
\hline
\textbf{Material Type}              & \textbf{Number of Layers} & \textbf{Thickness (mm)} \\ \hline
\multirow{4}{*}{\textbf{Cardboard}} & 1 layer                   & 2.801                   \\ \cline{2-3} 
                                    & 2 layers                  & 5.62                    \\ \cline{2-3} 
                                    & 3 layers                  & 9.183                   \\ \cline{2-3} 
                                    & 4 layers                  & 11.93                   \\ \hline
\multirow{4}{*}{\textbf{Plastic}}   & 1 layer                   & 0.44                    \\ \cline{2-3} 
                                    & 2 layers                  & 0.91                    \\ \cline{2-3} 
                                    & 3 layers                  & 1.53                    \\ \cline{2-3} 
                                    & 4 layers                  & 1.95                    \\ \hline
\multirow{4}{*}{\textbf{Mask}}      & 1 layer                   & 1.65                    \\ \cline{2-3} 
                                    & 2 layers                  & 3.23                    \\ \cline{2-3} 
                                    & 3 layers                  & 4.49                    \\ \cline{2-3} 
                                    & 4 layers                  & 6.13                    \\ \hline
\end{tabular}
}
\end{table}



\section{Thermal Dissipation Fingerprint for Multiple Objects} \label{sec:MIDASForMultipleObjects}

Our existing prototype and experiments are oriented to the individual characterization of objects using a single thermal dissipation fingerprint. In practice, humans tend to interact with many objects in their vicinity, and thermal dissipation fingerprints of multiple objects need to be analyzed simultaneously. This section describes an extension of the \systemname~pipeline to support the prediction of multiple objects and validates the extended pipeline through benchmark experiments.

\subsection{From Single to Multiple Objects}

\begin{figure}[!t]
	\centering	
	\includegraphics[width=1.0\columnwidth]{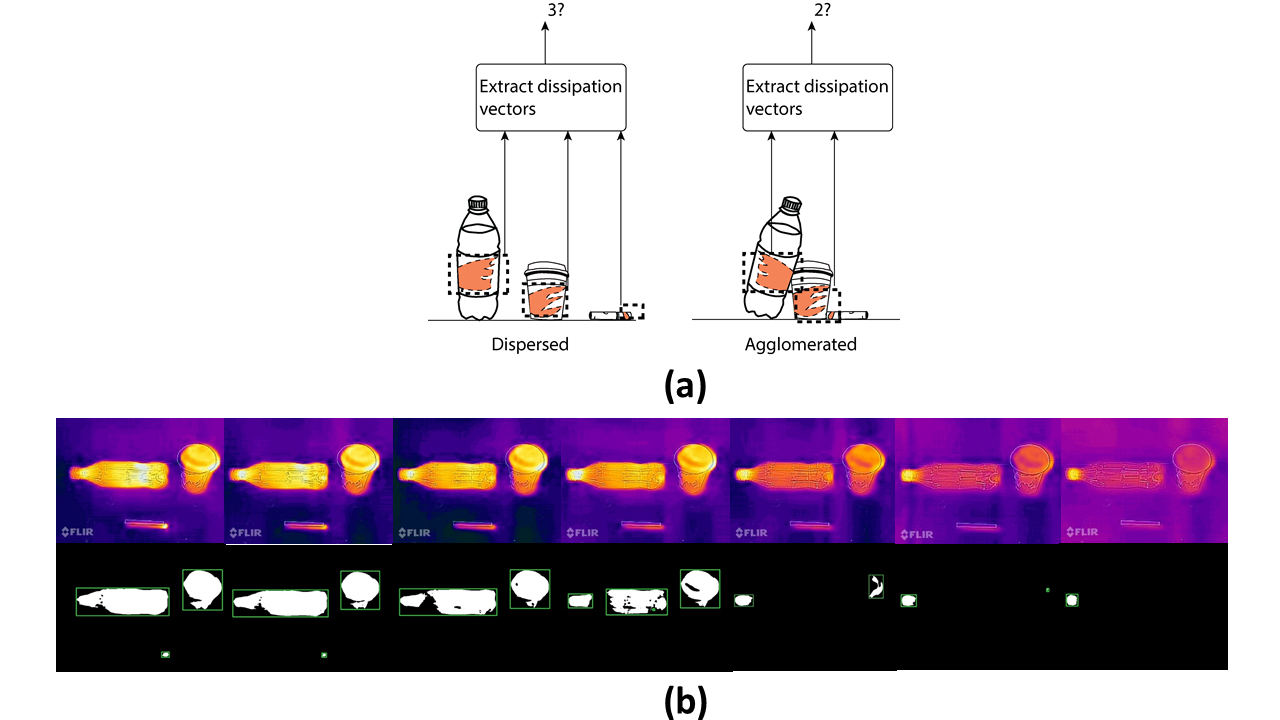}
	\caption{Detection of multiple thermal dissipation fingerprints. a) Object arrangements, b) Extraction of thermal dissipation vectors.}
	\label{fig:multipleobjectapproach}
	   \vspace{-.4cm}
\end{figure}

When analyzing the thermal fingerprints of multiple different objects, the main challenge is to separate and identify the different objects in the captured image. In thermal imaging, this task is significantly easier than in traditional computer vision. The dissipation of fingerprints is not sensitive to orientation or the shape of an object,  as they are easily distinguishable from the background. However, as shown in Figure~\ref{fig:multipleobjectapproach}a, object arrangement can play an important factor when extracting thermal dissipation vectors. The thermal fingerprint is easily recognizable individually when objects are not touching each other (dispersed), but it can spread over different surfaces if they touch each other (agglomerated).

We extend \systemname~to support multiple objects by including an additional step where the captured image is segmented into regions of interest (ROI)s. The current implementation operates on the normalized grayscale images and detects different ROIs using the~\textit{Counter Approximation Method}\footnote{Available in python CV2 package}; see  Figure~\ref{fig:multipleobjectapproach}b for an illustration. Once the ROIs are detected, each of them is mapped to a dissipation vector that corresponds to one object. As thermal radiation spreads and transfers through the surface of the objects, the identification needs to focus on the centroids of the hottest regions to ensure the ROIs correspond to different objects rather than transfer of dissipating heat along the surface of the object. 

\subsection{Evaluation}

We analyze the performance of our extended approach to recognize multiple objects simultaneously. We measure the degradation in response time when detecting multiple objects in the same images \rev{and we also evaluate the accuracy of the detection}. 


\cparagraph{Experimental Setup} The length (time) and frame rate of video footage that captures thermal dissipation are vital factors influencing the accurate identification of multiple objects. \rev{We conduct experiments where we fix the frame rate to $30$fps (i.e., $1800$ frames per minute) and conduct experiments on videos with different length:  (i) four videos of one-minute length; (ii) four videos from $30$ seconds of length to $120$ seconds in $30$ second increments.} In parallel, we also analyze the impact on performance as the number of objects to be identified increases. We analyze the identification from two to four objects simultaneously in different arrangements. Table~\ref{tab:grouping} describes the information on the grouping of object materials. We rely on the household objects used in our previous experiments. Specifically, we used a Coffee Cup (CC), Cigarette Butt (CB), Plastic Bottle (PB), and a Face Mask (mask).

\begin{table}[h]
\centering
\caption{Table explaining grouping arrangement for prediction of Multiple Objects}
\label{tab:grouping}
\begin{tabular}{@{}|c|l|@{}}
\toprule
\multicolumn{1}{|l|}{\textbf{Arrangement Group}} & \textbf{Description}         \\ \midrule
\textbf{A}                                       & \begin{tabular}[c]{@{}l@{}}One Object at a time. Coffee Cup \includegraphics[height=.5cm ]{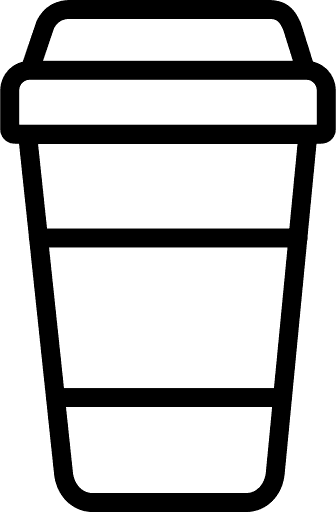}, \\ Cigarette Butt \includegraphics[height=.5cm ]{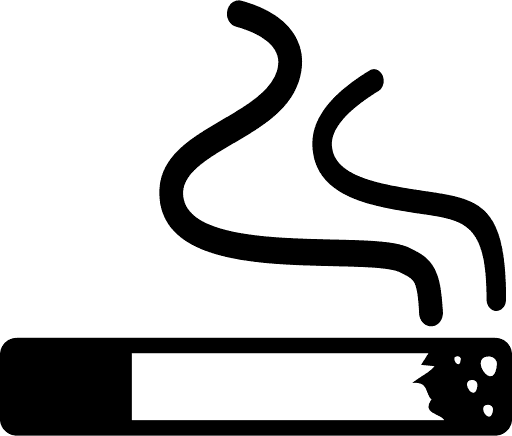}, \\ Plastic Bottle \includegraphics[height=.5cm ]{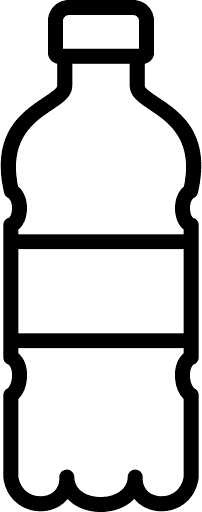}  \end{tabular}                                                 \\ \midrule
\textbf{B}                                       & \begin{tabular}[c]{@{}l@{}}Two Objects at a time. \\ "Coffee Cup + Plastic Bottle" \includegraphics[height=.5cm ]{graphics/cc.png} \includegraphics[height=.5cm ]{graphics/bottle.png}, \\ "Cigarette Butt + Plastic Bottle" \includegraphics[height=.5cm ]{graphics/cigg.png} \includegraphics[height=.5cm ]{graphics/bottle.png}, \\"Coffee Cup + Cigarette Butt" \includegraphics[height=.5cm ]{graphics/cc.png} \includegraphics[height=.5cm ]{graphics/cigg.png} \end{tabular} \\ \midrule
\textbf{C}                                       & \begin{tabular}[c]{@{}l@{}}Three Objects at a time. \\ "Coffee Cup, Plastic Bottle and Cigarette Butt"\\ \includegraphics[height=.5cm ]{graphics/cc.png} \includegraphics[height=.5cm ]{graphics/bottle.png} \includegraphics[height=.5cm ]{graphics/cigg.png}\end{tabular}                        \\ \midrule
\textbf{D}                                       & \begin{tabular}[c]{@{}l@{}}Four Objects at a time. "Coffee Cup, Plastic \\ Bottle, Cigarette Butt and Face Mask \\ \includegraphics[height=.5cm ]{graphics/cc.png} \includegraphics[height=.5cm ]{graphics/bottle.png} \includegraphics[height=.5cm ]{graphics/cigg.png} \includegraphics[height=.5cm ]{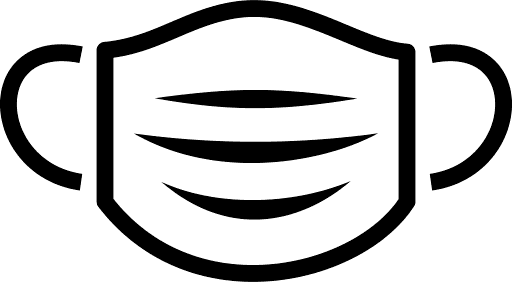} 
 \end{tabular}                   \\ \bottomrule
\end{tabular}
\end{table}

\begin{figure}
	\centering
		\includegraphics[width=0.75\columnwidth]{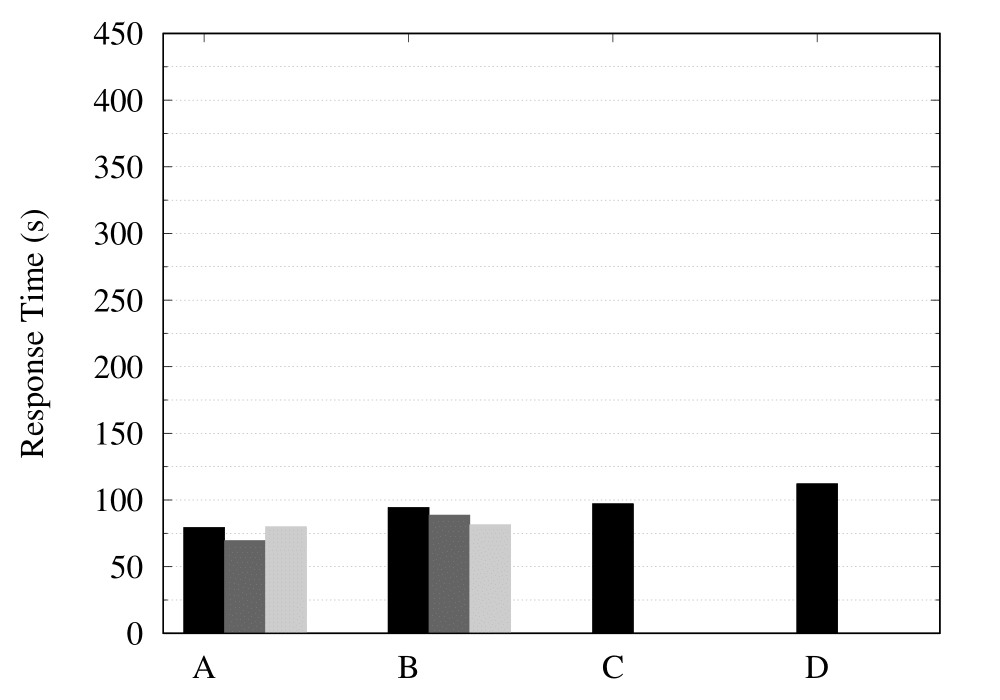}
	\caption{Response time of multiple object identification when using \systemname. (same video configuration). \rev{The bars within the figure correspond to different individual objects (A) or combination of objects (two objects: B, three objects: C, four objects: D) as given in Table~\ref{tab:grouping}.}}
	\label{figure:equallengthvideo}
\end{figure}

\cparagraph{Results - Dispersed case} Figure~\ref{figure:equallengthvideo} shows the results of our first experiment where videos of the same length and frame rate are used. 
\rev{The processing time of the system naturally increases as more dissipation vectors are given as input but the overall increase is negligible. In terms of arrangement of objects, the results similarly are stable, suggesting that the response time of \SystemName{} is not affected by the arrangement of objects -- provided that the dissipation vectors of the materials that form the arrangement can be extracted in a noise-free fashion.}

\begin{figure}[!t]
\centering
    \begin{subfigure}[b]{.45\linewidth}
    \centering
    \includegraphics[width=1.0\textwidth]{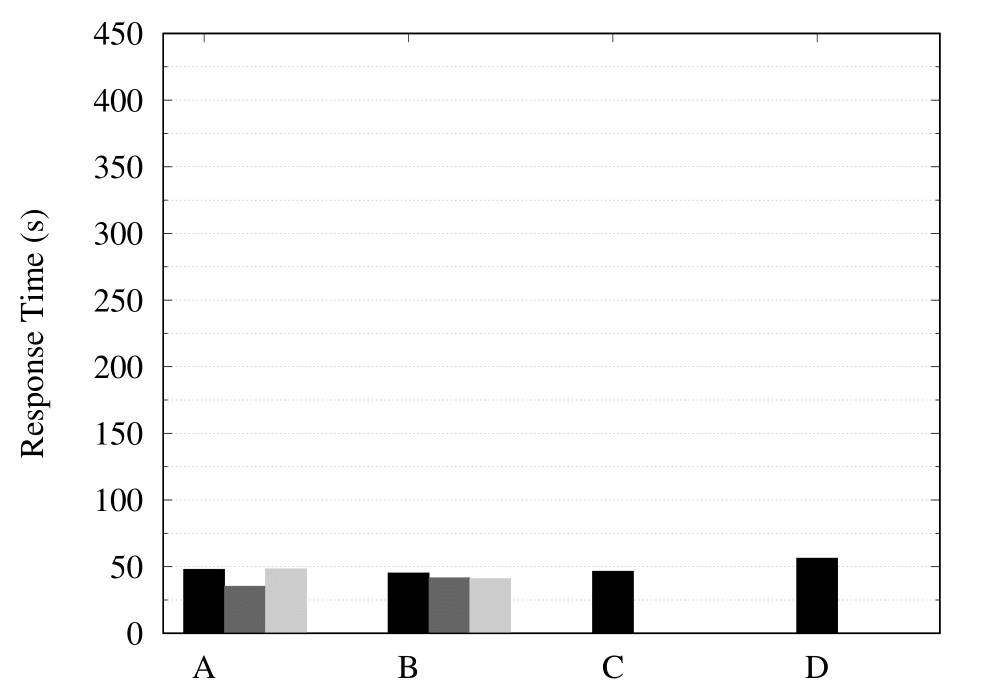}
   \caption{30 second video (900 frames)}\label{fig:30secsvideo}
  \end{subfigure}   
  \begin{subfigure}[b]{.45\linewidth}
    \centering
    \includegraphics[width=1.0\textwidth]{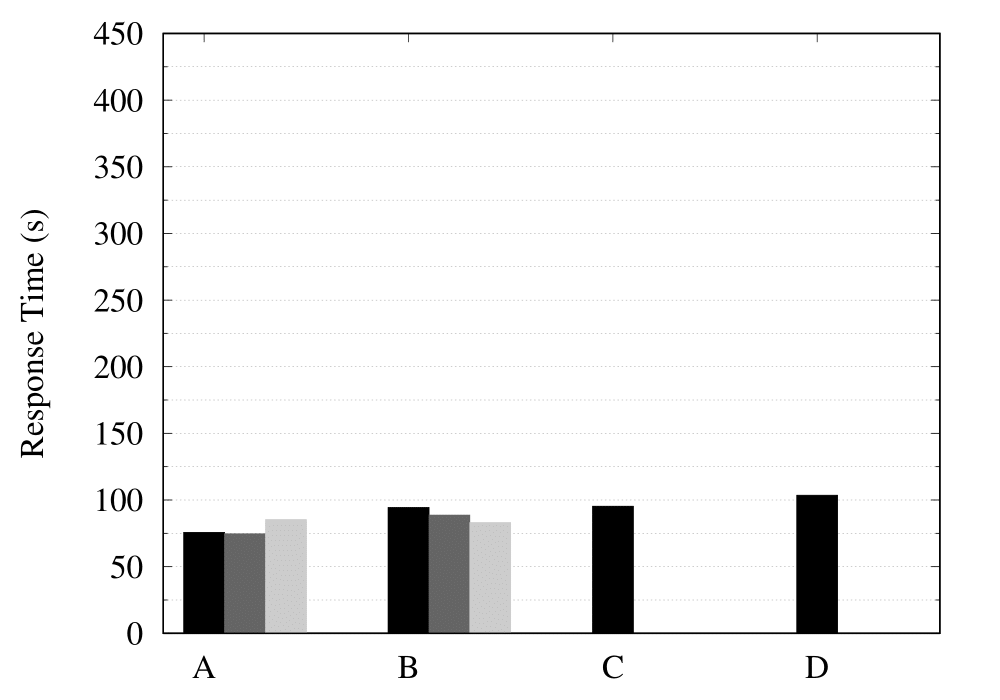}
    \caption{1 minute video (1800 frames)}\label{fig:1minvideo}
  \end{subfigure} 
  \begin{subfigure}[b]{.45\linewidth}
    \centering
    \includegraphics[width=1.0\textwidth]{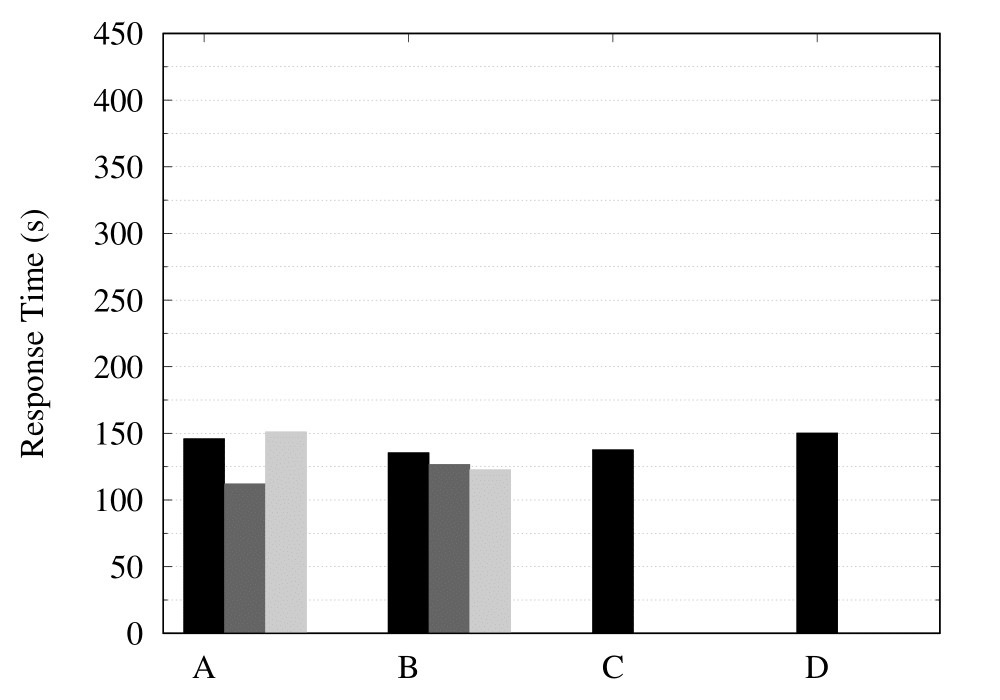}
    \caption{1 minute 30 second video (2700 frames)}\label{fig:1min30secsvideo}
  \end{subfigure} 
  \begin{subfigure}[b]{.45\linewidth}
    \centering
    \includegraphics[width=1.0\textwidth]{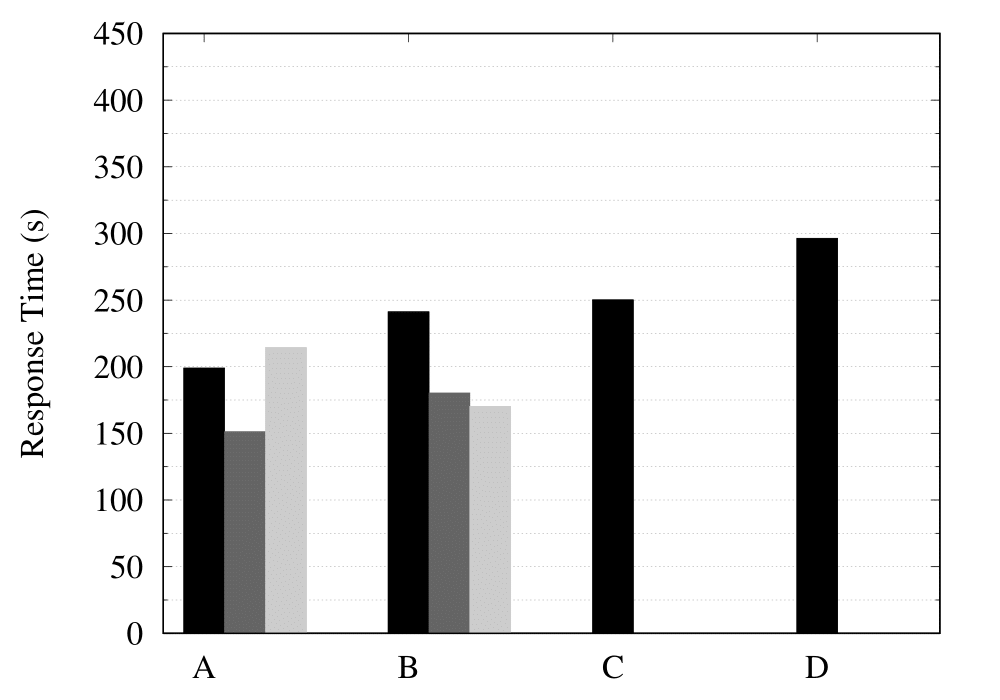}
    \caption{2 minute video (3600 frames)}\label{fig:2minvideo}
  \end{subfigure} 
  \caption{Impact on response time when using videos of different length and frame rate (Disperse objects case). \rev{The bars within the figure correspond to different individual objects (A) or combination of objects (two objects: B, three objects: C, four objects: D) as given in Table~\ref{tab:grouping}.}} 
  \label{fig:variablelengthvideo}
	   \vspace{-.4cm}
\end{figure}

Figure~\ref{fig:variablelengthvideo} shows the results for the case where different video configurations are considered. \rev{As the length of the video feed increases, the response time of \SystemName{} naturally increases as well. For videos up to $90$ seconds, the length of the video dominates the performance and only once the video length reaches $120$ seconds the impact of segmenting multiple objects becomes visible. In practice the thermal dissipation fingerprints are stable across the video lengths and hence using a short video length of $30$ seconds is sufficient for most situations.} 
\rev{The results also show that the response time does not always increase. This is because the response time tends to be dominated by the longest dissipation time and thus cases where one object has significantly longer dissipation time than others tend to have fairly constant response time. Indeed, the response time depends on the calculation of the area reduction in the image. Once the thermal fingerprint has dissipated, the area is effectively zero and the computations return fast. In the experiments, both the plastic bottle and coffee cup have longer dissipation fingerprints than the cigarette butt and thus the response time leans toward the longest vectors.}

\begin{figure}[!t]
\centering
    \begin{subfigure}[b]{.45\linewidth}
    \centering
    \includegraphics[width=1.0\textwidth]{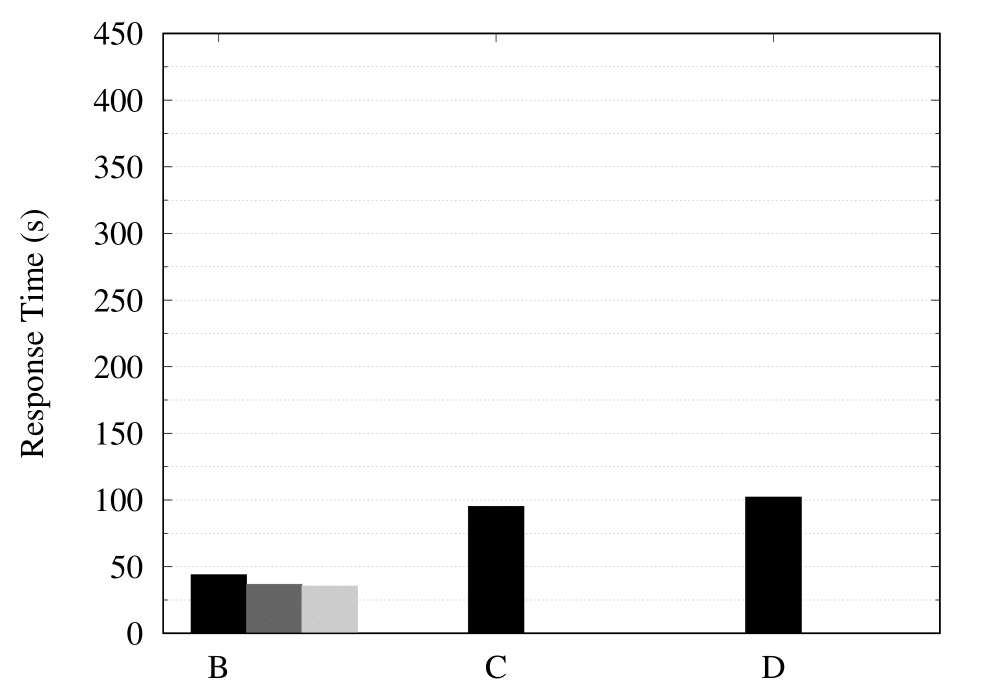}
   \caption{30 second video (900 frames)}\label{fig:30secsvideoagglomerated}
  \end{subfigure}   
  \begin{subfigure}[b]{.45\linewidth}
    \centering
    \includegraphics[width=1.0\textwidth]{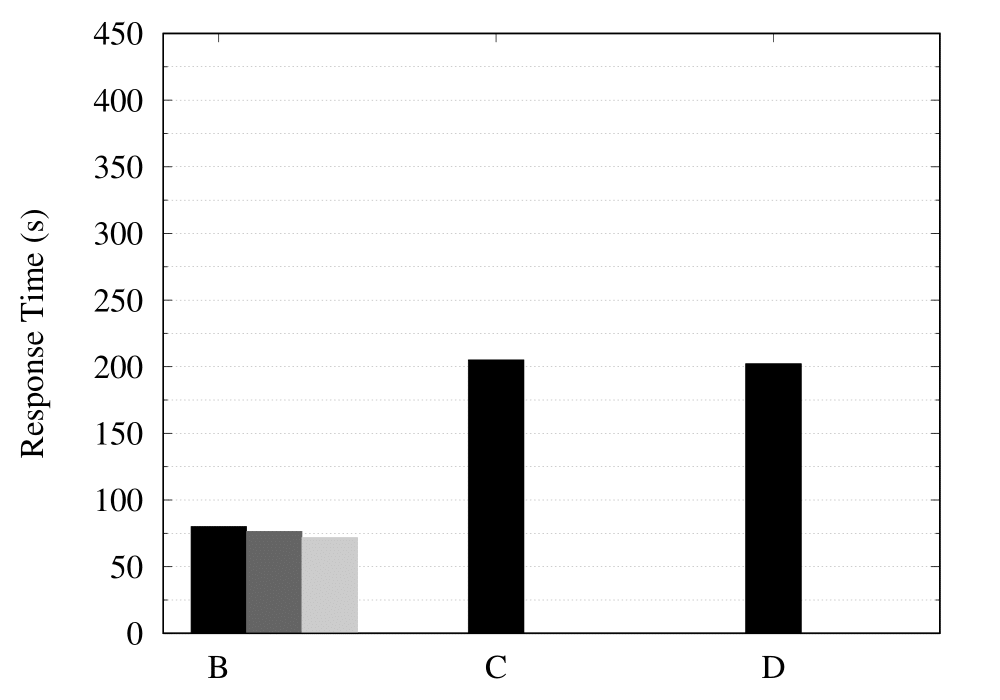}
    \caption{1 minute video (1800 frames)}\label{fig:1minvideoagglomerated}
  \end{subfigure} 
  \begin{subfigure}[b]{.45\linewidth}
    \centering
    \includegraphics[width=1.0\textwidth]{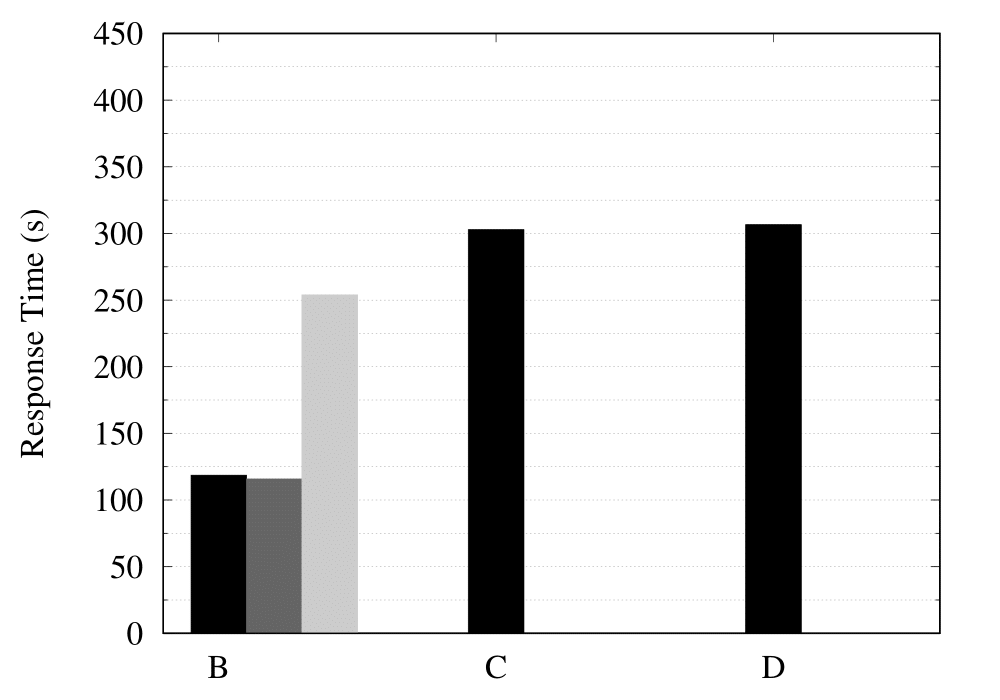}
    \caption{1 minute 30 second video (2700 frames)}\label{fig:1min30secsvideoagglomerated}
  \end{subfigure} 
  \begin{subfigure}[b]{.45\linewidth}
    \centering
    \includegraphics[width=1.0\textwidth]{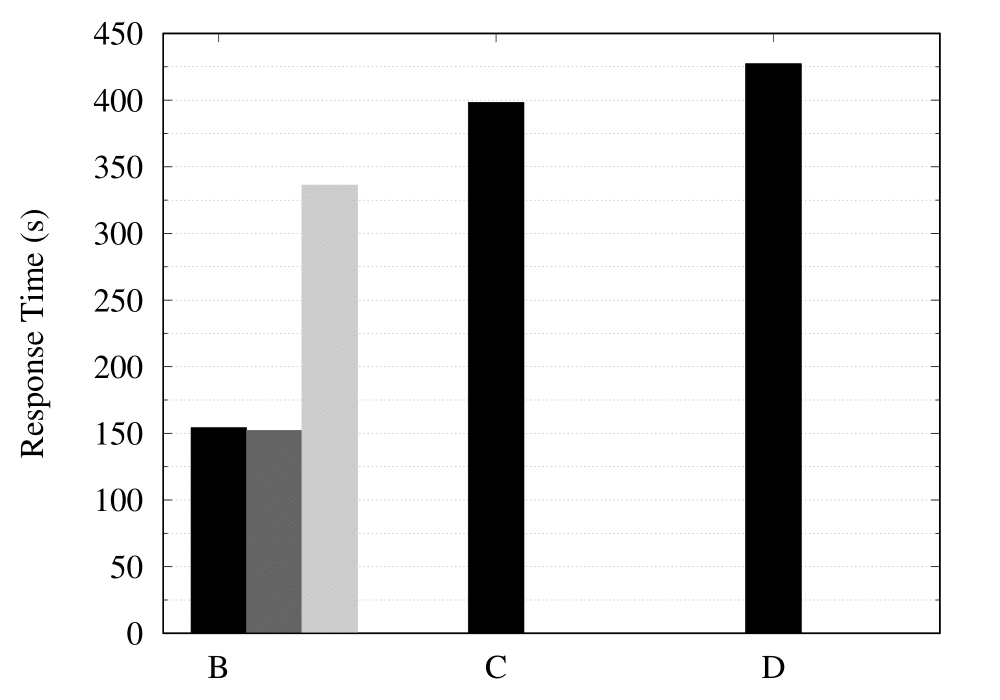}
    \caption{2 minute video (3600 frames)}\label{fig:2minvideoagglomerated}
  \end{subfigure} 
  \caption{Impact on response time when using videos of different length and frame rate (Agglomerated objects case). \rev{The bars within the figure correspond to different combinations of objects (two objects: B, three objects: C, four objects: D) as given in Table~\ref{tab:grouping}.}} 
  \label{fig:variablelengthvideoagglomerated}
	   \vspace{-.4cm}
\end{figure}

\cparagraph{Results - Agglomerated case} 
\rev{When the objects or the areas of their thermal dissipation area intersect, this can result in dissipation vectors becoming mixed. We next investigate the performance of \systemname{} in such cases.}  
Figure~\ref{fig:variablelengthvideoagglomerated} shows the processing time of \SystemName{} when materials are in contact. As shown in the figure, we have only considered arrangement groups B,C, and D as we are evaluating multiple objects that are in contact. From the result, we can observe higher computational overhead, which is caused mostly due to the continuous separation of thermal dissipation vectors to characterize individual objects. 

\cparagraph{Effect on Accuracy} \rev{We next assess the effect multiple objects have on the accuracy of \SystemName{}. We only consider the dispersed case as the performance depends on the quality of the segmentation, which is difficult to control systematically. Indeed, in rare cases the segmentation may fail when the thermal fingerprints of two objects appear as a single "blob" in the image, making it impossible to separate them. When the segmentation fails, naturally the following recognition fails.}

\begin{figure}
	\centering
		\includegraphics[width=0.75\columnwidth]{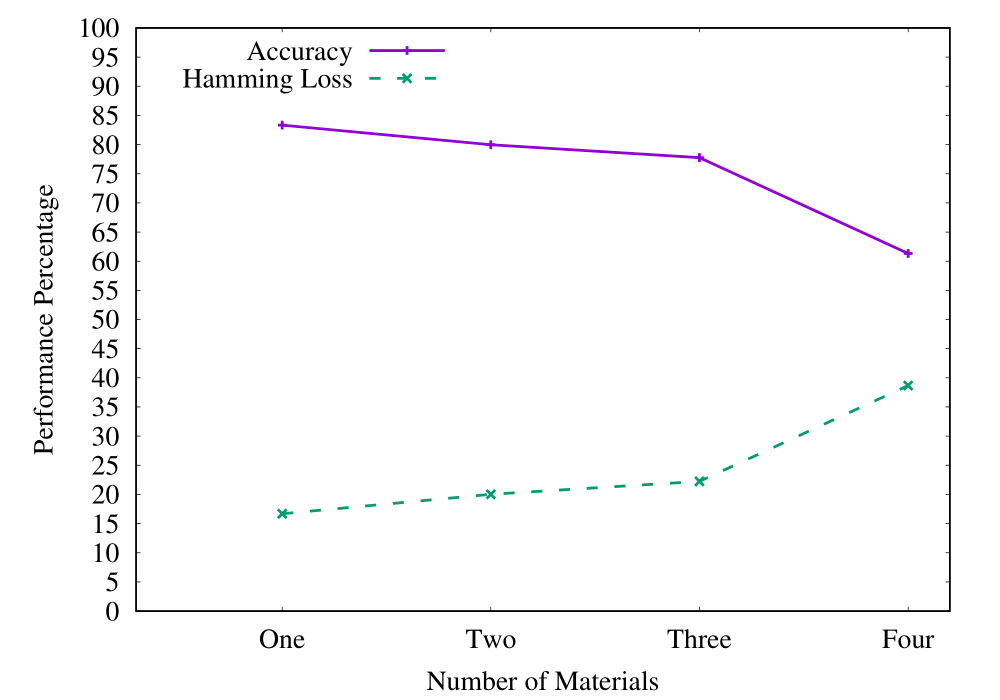}
	\caption{Average prediction accuracy for different number of objects and the corresponding Hamming loss.}
	\label{fig:accuracy}
\end{figure}

\rev{The detection accuracy for multiple object detection is shown in Figure~\ref{fig:accuracy}. For predicting single material objects, the accuracy is $83$\%. For evaluating the effect on accuracy with multiple materials concurrently, we start with combinations based on the arrangement groups from Table~\ref{tab:grouping}. We feed the model with combinations from the different arrangement groups with five different input video footage but involving same combinations to evaluate the average prediction accuracy with multiple materials. We then compare the predictions labels with the true labels and note the average accuracy. For two materials based on arrangement B, the average accuracy achieved was $79.97$\%. For arrangement C with three materials, the average accuracy is $77.75$\% and finally for the arrangement D with four objects, the average accuracy is $61.33$\%. The performance overall depends on the quality of the segmentation, the resolution of the image, and the type of object in the image. As more objects enter the field-of-view, the size of the objects is necessarily smaller than in the case where only a single object is visible. This reduces the quality of the segmentation and the dissipation area, and degrades overall classification performance. Similarly, the four object case is the only configuration that considers the face mask and part of the performance decrease results from difficulty in recognizing the mask. Note that in this type of cases, the objects resulting in poor detection can be separately flagged and they can be analyzed individually to improve the quality of the analysis.}

\section{Discussion}
\label{sec:Discussion}

\cparagraph{Human temperature} Human temperature changes in cycles, being at its highest during hours of activity (day) and lowest during sleep (night)~\cite{obermeyer2017individual}. We demonstrate that interactions with objects can be used to characterize materials. The best results are obtained when the body temperature is stable, but the relative differences in thermal fingerprints are consistent across variations in body temperature. Conversely, interactions with an object of known material and in a stable environment can be used to detect relative differences in body temperature.


\cparagraph{Room for improvement} \rev{Naturally, there are further challenges that need to be addressed to make our solution more robust across diverse environments.} Adapting our approach to continuous monitoring requires accurate and noise-free thermal images, e.g., using calibration~\cite{malmivirta2019hot}. Not all materials can be characterized using our solution as thermal cameras have different emissivity ranges and some materials may reflect too much -- or too little -- thermal radiation. Such materials are usually used to preserve user's privacy, e.g., ATM pin codes~\cite{abdelrahman2017stay}. The experiments were conducted using thermal cameras integrated into smartphones, but in the future it may be possible to use cheaper alternatives, such as low-cost thermal array sensors~\cite{rintahomi21how}. \rev{There are also challenges arising from practical deployments as environments with oscillating temperatures can result in unstable fingerprints and as there can be situations where only partial dissipation fingerprints are available (e.g., due to the use of gloves).}  While outside the scope of our present work, there is also room for developing application areas, e.g., validating temperature differences with patients as part of clinical studies.  


\cparagraph{Other material properties} Dissipation time of thermal fingerprint gives insights about material types and correlates with emissivity. Thermal imaging could be used to potentially infer other material properties, such as thickness and elasticity. Potential use cases include detecting the pollutant type of marine plastics~\cite{flores2021toward} and monitoring the decay in organic materials using differences in thermal dissipation characteristics. 

\cparagraph{Micro-expressions through hand-touch} Micro-expressions are facial expressions captured in a fraction of time by cameras. Our work demonstrates that the thermal fingerprint of a hand-touch on an object has a life span that depends on how long the object was held. Thus, thermal analysis of human touch can be another approach for detecting micro-expressions, e.g., disgust and excitement. Similarly, micro-expressions through human-touch can be envisioned as an approach to measuring individuals' productivity at work. 

\cparagraph{Robots and autonomous devices} Thermal radiation analysis of objects touched by humans can be used to inform and train different robots and autonomous devices, e.g., UAVs~\cite{liyanage2021geese}, about the material properties of objects. New sensing and interaction modalities can also be envisioned as part of robotic systems, e.g., incorporating heat sensation to detect the material of an object and to enable autonomous devices to adjust their operations with objects in the surrounding environment~\cite{li2020skin}. For instance, a robotic arm can rely on a camera to detect an empty bottle on a table. However,  the arm's pressure to lift and put the bottle should be proportional to the bottle material. Otherwise, the robotic arm can break apart the bottle, e.g., plastic vs. glass bottle. Besides, since it is possible to differentiate thermal radiation emitted by different genders, autonomous devices can adjust their interactions accordingly. 


\cparagraph{Augmented Reality systems} Augmented Reality systems that mix the real and virtual worlds can benefit from thermal radiation monitoring. By piggybacking the human-touch on objects, it is possible to recognize the materials of objects further. These objects are mapped to the virtual world by considering their inherent natural material properties, such as wood, concrete, glass, and plastic.



\cparagraph{Application scenarios} \rev{The core focus of our work has been on showing the potential of using thermal dissipation as a sensing modality without focusing on any specific application scenario. Nevertheless, thermal dissipation has significant potential for many real-world application scenarios. One application area is household waste sorting as humans interact with waste objects just before they are thrown to the bin. Our approach allows using the residual thermal energy resulting from these interactions to identify which type of material the object consists of and to support better recycling practices. In our ongoing work, we have also shown how thermal dissipation resulting from sunlight can be integrated with autonomous ground vehicles (AGVs) to classify litter thrown on the ground~\cite{yin2022toward}. Our results also showed that thermal dissipation can be sensed from the inside of an object and this could be used as an alternative to identify interactions. For example, an interactive toy could detect when a child plays with it~\cite{fan2014injecting}. Another interesting application would be to integrate the sensors as part of medical bottles to support medication management~\cite{klakegg2018assisted}. Exploring these opportunities is an interesting venue for future research. Naturally, realising these scenarios requires further advances in thermal sensing technology. Firstly,  miniaturization of thermal cameras is required. While currently USB size solutions are available, they generally expect a sufficient power source and processing unit (e.g., a smartphone). It is expected further miniaturization of the technology as the one achieved by existing RGB-camera, e.g., pen-size cameras. Secondly, thermal dissipation time also needs to be characterized for this purpose. Indeed, in our experiments, we demonstrate that while thermal dissipation time can be measured from both sides of the object (inner and outer sides), both thermal dissipation differed as a result of thermal loss. Thus, thermal dissipation time has to be pre-computed considering different thicknesses. }

\section{Related Work}

\cparagraph{Thermal imaging}
The usage of thermal imaging has been studied in different domains and applications with examples ranging from monitoring the manufacturing process of smartphone hardware components~\cite{xie2014therminator} to medical analysis~\cite{sun2011portable,harangi2011detecting}. Other examples include facial recognition for bio-metric authentication~\cite{bhowmik2008classification}, cognitive analysis~\cite{abdelrahman17cognitive}, gestures~\cite{larson2011heatwave, abdelrahman2015investigation}, and energy modeling of IoT devices~\cite{flores2019evaluating}. Our work extends thermal imaging to material classification. 



\cparagraph{Material sensing} Materials have different characteristics different properties that can be exploited to categorize them. The most common material sensing approach is to rely on different parts of the light spectrum and measure either reflection or absorption at different frequencies. Examples range from the use of green light sensing to detect plastic waste~\cite{flores2020penguin} to the use of near-infrared sensing to facilitate medicine adherence~\cite{klakegg2018assisted} and the use of hyperspectral imaging for estimating sugar content in drinks~\cite{jiang2019probing}. Also, deep learning approaches for detecting different material types from reflection patterns at different wavelengths have been proposed~\cite{cho2018deep}. 
Our work extends these by using thermal radiation in the infrared spectrum to estimate internal characteristics of materials through heat dissipation.

\cparagraph{Sensorless sensing} 
Wireless signals can also be used to identify properties in materials. Examples include the use of variations in WiFi signal propagation characteristics to identify liquids~\cite{dhekne2018liquid}, and the use of surface tension to characterize liquids~\cite{yue2019liquid,wei2015study}. These methods generally require either close contact with the material or a transmitter - receiver pair to be placed on opposite sides of the material. Our work offers a non-contact technique for material characterisation that piggybacks thermal radiation generated from humans. 

\section{Summary and Conclusions}


We developed \SystemName{} as an innovative sensing solution for characterizing and recognizing everyday objects from the dissipation of residual thermal energy resulting from human touch. \SystemName{} uses thermal imaging to monitor thermal energy changes over time and models these changes to infer and characterize the material type of an object. Through extensive empirical benchmarks, we demonstrated that changes in thermal fingerprints are robust to variations in the way people interact with objects and the people themselves. \SystemName{} recognizes different material with up to $83\%$ accuracy using only the dissipation of thermal fingerprints. Our solution offers an innovative sensing solution for classifying materials and taking advantage of human interactions with everyday objects.

\section{Acknowledgments}
This research is supported by the European Social Fund via the IT Academy Programme and the Academy of Finland grant 339614. The publication only reflects the authors' views. The authors thank the anonymous reviewers for their insightful comments.

\section{References}

\bibliographystyle{elsarticle-num}
\bibliography{MIDAS-MEC}

\end{document}